\documentclass[12pt]{article}
\usepackage{amsmath,lscape,amsfonts,amssymb,amsthm,verbatim,color,lscape}
\usepackage[figuresright]{rotating}
\usepackage{natbib}
\usepackage{url}
\usepackage{hyperref}
\hypersetup{colorlinks,
citecolor=black,
filecolor=black,
linkcolor=black,
urlcolor=black,
pdftex}

\usepackage{anysize}
\marginsize{2cm}{2cm}{2cm}{2cm}

\usepackage{colortbl}
\definecolor{gray}{rgb}{0.8,0.8,0.8}

\usepackage{times}
\usepackage{blindtext}
\usepackage[font={footnotesize,it}]{caption}

\usepackage{sectsty}  
\subsectionfont{\normalsize\bf}
\sectionfont{\large\bf}

\def \red#1{\textcolor{red}{#1}}

\def \red#1{\textcolor{red}{#1}}

\begin{document}

\def\p{\partial}
\def\oo{\infty}
\def\rt#1{\sqrt{#1}\,}

\def\Cbar{{\overline C}}
\def\C{\mathbf{C}}
\def\E{{\rm E}\,}
\def\I{\mathbf{I}}
\def\pp{\mathbf{p}}
\def\R{\mathbf{R}}
\def\y{\mathbf{y}}
\def\Y{\mathbf{Y}}
\def\z{\mathbf{z}}
\def\x{\mathbf{x}}
\def\o{\omega}
\def\s{\sigma}

\def\V{\mathbf{V}}
\def\I{\mathbf{I}}
\def\bfv{\mathbf{v}}
\def\X{\mathbf{X}}
\def\D{\mathbf{D}}

\def\a{\boldsymbol{\alpha}}
\def\g{\gamma}
\def\b{\beta}

\def\de{\delta}
\def\debf{\boldsymbol{\delta}}
\def\e{\epsilon}
\def\th{\theta}
\def\thbf{\boldsymbol{\theta}}
\def\pibf{\boldsymbol{\pi}}
\def\Xibf{\boldsymbol{\Xi}}
\def\Sbf{\boldsymbol{\Sigma}}

\def \red#1{\textcolor{red}{#1}}
\def \blue#1{\textcolor{blue}{#1}}
\def \magenta#1{\textcolor{magenta}{#1}}
\def \green#1{\textcolor{green}{#1}}
\def\bbf{\boldsymbol{\beta}}
\def\bmu{\boldsymbol{\mu}}

\long\def\symbolfootnote[#1]#2{\begingroup
\def\thefootnote{\fnsymbol{footnote}}\footnote[#1]{#2}\endgroup}
\newcommand{\strike}{\color{red}\sout}

\newcommand{\widesim}[2][1.5]{
  \mathrel{\overset{#2}{\scalebox{#1}[1]{$\sim$}}}
}

\title{Hybrid copula mixed models for combining case-control and cohort studies in meta-analysis  of diagnostic tests}

\date{}
\author{
Aristidis K. Nikoloulopoulos\footnote{{\small\texttt{A.Nikoloulopoulos@uea.ac.uk}}, School of Computing Sciences, University of East Anglia,
Norwich NR4 7TJ, UK}
}
\maketitle

\vspace{2ex}

\begin{abstract}
\baselineskip=20pt
\noindent
Copula mixed models for trivariate (or bivariate) meta-analysis of diagnostic test accuracy studies accounting (or not) for disease prevalence have been proposed in the biostatistics literature to synthesize information. However, 
many systematic reviews  often include case-control and cohort studies, so one can either focus on the bivariate meta-analysis of the case control studies or the trivariate meta-analysis of the cohort studies, as only the latter contains information on disease prevalence. In order to remedy  this situation of wasting data we propose a hybrid copula mixed  model  via a combination of the bivariate and trivariate copula mixed model for the data from the case-control studies and   cohort studies, respectively. Hence, this hybrid model can account for study design and also due its generality can  deal with dependence in the joint tails. We apply the proposed hybrid copula mixed model to a review of the performance of contemporary diagnostic imaging modalities for detecting metastases in patients with melanoma.
\end{abstract}

\maketitle

\baselineskip=16pt

\section{Motivating study and background}

Melanoma is the least common but most deadly type of skin cancer and occurs in melanocytes, which are cells that produce the skin pigment melanin \citep{Jerant-etal-2000}.
A systematic review of published studies  by \cite{Xing-etal-2011} has examined the accuracy of contemporary diagnostic imaging modalities for detecting metastases in patients with melanoma and identified 60 cohort and 43 case-control studies.  

\cite{Xing-etal-2011} applied the generalized linear mixed model (GLMM), proposed by  \cite{Chu&Cole2006}, to account for the association between the sensitivity and specificity across studies. 
However, it is reported in the literature that the assumption of independence  between the  sensitivity/specificity with disease prevalence in the bivariate GLMM is likely to be violated
\citep{brenner-gefeller-1997,leeflang-etal-2009,Leeflang-etal-2013}.  By fitting the bivariate GLMM 
the information on prevalence of melanoma, which is available only in cohort studies, has been totally neglected, and,  thus an important  amount of data  has been wasted.

\cite{chu-etal-2009}   extended the bivariate GLMM to a trivariate GLMM by also accounting for disease prevalence.
Nevertheless, this model  can only meta-analyse data from the   cohort studies,  since the disease prevalence is not available in case-control studies.   Very recently, \cite{chen-etal-2015-jrssc} developed a hybrid model that exploits the use of both the bivariate and trivariate GLMM for combining case-control and cohort studies (hereafter hybrid GLMM) and applied the model to fully analyse the systematic review of published studies  in \cite{Xing-etal-2011}  Due to the fact that they noticed computational problems such as non-convergence and singularities,   they developed a composite likelihood (CL) method to overcome the computational difficulties on the estimation of the hybrid GLMM. The CL method is well established in the statistical literature as a surrogate alternative of maximum likelihood (ML) when the joint likelihood is too difficult to compute \citep{varin08,Varin-etal2011}. The advantage of the CL approach in this application domain  is that the likelihood can  be derived conveniently under the assumption of independence between the random effects, i.e., the  latent vector of transformed sensitivity, specificity, and disease prevalence.  
\cite{Chen-etal-smmr-2014}
proposed a CL method  even for the  estimation of  the GLMM to overcome practical `issues' in the joint likelihood inference such as  computational difficulty caused by a double integral in the joint likelihood function. Our view is that 
GLMM can only be unstable if there are too many parameters in the covariance
matrix of the random effects or too many random effects for a small
sample \citep{Demidenko04}, which is not the case in this application domain.

\cite{Nikoloulopoulos2015b,
Nikoloulopoulos2015c}
proposed  copula mixed models for bivariate and trivariate   meta-analysis of diagnostic test accuracy studies and made the argument for moving to the general class of copula random effects models.
The copula mixed models include the bivariate and trivariate GLMMs \citep{Chu&Cole2006,chu-etal-2009} 
 as special cases,  can also operate on the original scale of sensitivity,  specificity, and disease prevalence, and their estimation can be successfully approached by ML estimation. 
 
In this paper building in the aforementioned papers, we propose a hybrid copula mixed model to combine case-control and cohort studies. We combine the bivariate and trivariate copula mixed model for the data from the case-control studies and cohort studies, respectively. 
The hybrid copula mixed has as special case the hybrid GLMM and  features several other  advantages:  (a) the random effects distributions are expressed via copulas which allow for flexible dependence modelling, different from assuming simple linear correlation structures, normality and tail independence  (b)  can also operate on the original scale of sensitivity, specificity, and prevalence, and (c) estimation can be approached  by the `gold standard'  ML method.

The remainder of the paper proceeds as follows. Section \ref{sec-model} introduces  the  hybrid copula mixed model for diagnostic test accuracy (case-control and cohort) studies. 
An ML estimation technique and computational details are provided in  Section \ref{sec-est}.     Section \ref{sec-sim}  contains  small-sample  efficiency calculations
to  investigate the effect of misspecifying the random effects distributions  and compare the proposed methodology to the CL approach proposed by \cite{chen-etal-2015-jrssc}.
In Section \ref{sec-app} we analyse   the  systematic review of the accuracy of contemporary diagnostic imaging modalities for detecting metastases in patients with melanoma and show   efficiency gains with respect to the CL approach.
We conclude with some discussion in Section \ref{sec-disc}.

\section{\label{sec-model}The hybrid copula mixed model}

In this section we introduce the hybrid copula mixed model. Before that we provide some background about important tools to form the hybrid copula mixed model. These are a brief introduction  to copulas in Subsection \ref{overview}, the bivariate copula mixed model in Subsection \ref{2model}, and the  vine copula mixed model in Subsection \ref{3model}.

\subsection{\label{overview}Overview and relevant background for copulas}
A copula is a multivariate cdf with uniform $U(0,1)$ margins \citep{joe97,joe2014,nelsen06}.
If $F$ is a $d$-variate cdf with univariate margins $F_1,\ldots,F_d$,
then Sklar's (1959)\nocite{sklar1959}  theorem implies that there is a copula $C$ such that
  $$F(x_1,\ldots,x_d)= C\Bigl(F_1(x_1),\ldots,F_d(x_d)\Bigr).$$
The copula is unique if $F_1,\ldots,F_d$ are continuous.
If $F$ is continuous and $(Y_1,\ldots,Y_d)\sim F$, then the unique copula
is the distribution of $(U_1,\ldots,U_d)=\left(F_1(Y_1),\ldots,F_d(Y_d)\right)$ leading to
  $$C(u_1,\ldots,u_d)=F\Bigl(F_1^{-1}(u_1),\ldots,F_d^{-1}(u_d)\Bigr),
  \quad 0\le u_j\le 1, j=1,\ldots,d,$$
where $F_j^{-1}$ are inverse cdfs \citep{nikoloulopoulos&joe12}.  For example,
if $\Phi_d(\cdot;\R)$
is the MVN cdf with correlation matrix $$\R=(\rho_{jk}: 1\le j<k\le d)$$ and
N(0,1) margins, and $\Phi$ is the univariate standard normal cdf,
then the MVN copula is
\begin{equation}\label{MVNcdf}
C(u_1,\ldots,u_d)=\Phi_d\Bigl(\Phi^{-1}(u_1),\ldots,\Phi^{-1}(u_d);\R\Bigr).
\end{equation}

In the bivariate case  there are many parametric families of copulas.  
However, 
their multivariate extensions have limited dependence structures. An approach to successfully subside this restriction is the vine pair-copula construction \citep{Kurowicka-Joe-2011,joe2014} 
which is based on $d(d-1)/2$ bivariate copulas, of which some are
used to summarize conditional dependence.
Vine copulas include the MVN as special case, but can also cover reflection
asymmetry and have upper/lower tail dependence parameters being different
for each bivariate margin \citep{joeetal10}.
Vines require a decision on the indexing of variables.  For example, for a 3-dimensional vine  copula there are $3$ distinct permutations: $$\{12,13,23|1\}, \qquad \{12,23,13|2\}, \quad \mbox{and} \quad \{13,23,12|3\}.$$ 
For each of them, the 3-dimensional vine  is decomposed on $3$ bivariate copulas, of which the one is
used to summarize conditional dependence; see \cite{Nikoloulopoulos2015c} for more details.

Table \ref{2fam} provides   a sufficient list of bivariate copulas  for meta-analysis of diagnostic test accuracy studies      \citep{Nikoloulopoulos2015b,
Nikoloulopoulos2015c}.
These copula families have different strengths of tail behaviour and tail dependence is a property to consider when choosing amongst different families of copulas and the concept of upper/lower tail dependence is one way to differentiate families.  \cite{Nikoloulopoulos&karlis08CSDA} have shown that it is hard to choose a copula with similar properties from real data, since copulas with similar (tail) dependence properties provide similar fit.

\setlength{\tabcolsep}{3pt}
\begin{table}[!h]
\caption{\label{2fam}Parametric families of bivariate copulas and their Kendall's $\tau$ as a strictly increasing function of the copula parameter $\theta$.}
\begin{small}
\centering
\begin{tabular}{ccc}
\hline
Copula & $C^{-1}(v|u;\th)$& 
$\tau$\\\hline
BVN & $\Phi\Bigl(\sqrt{1-\th^2}\Phi^{-1}(v)+\th\Phi^{-1}(u)\Bigr)$ 
&$\frac{2}{\pi}\arcsin(\th)\quad ,\quad -1\leq\th\leq1$\\
Frank &$
-\frac{1}{\theta}\log\left[1-\frac{1-e^{-\th}}{(v^{-1}-1)e^{-\th u}+1}\right]
$
&$\begin{array}{ccc}
1-4\theta^{-1}-4\theta^{-2}\int_\theta^0\frac{t}{e^t-1}dt &,& \th<0\\
1-4\theta^{-1}+4\theta^{-2}\int^\theta_0\frac{t}{e^t-1}dt &,& \th>0\\
\end{array}$\\
Clayton  &$\Bigl\{(v^{-\theta/(1+\theta)}-1)u^{-\th}+1\Bigr\}^{-1/\theta}$
 &$\th/(\th+2)\quad ,\quad \th>0$\\
Clayton by 90 &$\Bigl\{(v^{-\theta/(1+\theta)}-1)(1-u)^{-\th}+1\Bigr\}^{-1/\theta}$&
$-\th/(\th+2)\quad ,\quad \th>0$\\
Clayton by 180 &$1-\Bigl[\bigl\{(1-v)^{-\theta/(1+\theta)}-1\bigr\}(1-u)^{-\th}+1\Bigr]^{-1/\theta}$
&$\th/(\th+2)\quad ,\quad \th>0$\\
Claytonby 270 &$1-
\Bigl[\bigl\{(1-v)^{-\theta/(1+\theta)}-1\bigr\}u^{-\th}+1\Bigr]^{-1/\theta}$
&$-\th/(\th+2)\quad ,\quad \th>0$\\
\hline

\end{tabular}
\end{small}
\end{table}

\subsection{\label{2model}Bivariate copula mixed model}

For each study $i$, the within-study model assumes that the number of true positives $Y_{i1}$ and true negatives $Y_{i2}$ are conditionally independent and binomially distributed given $\X=\x$, where $\X=(X_1,X_2)$ denotes the  bivariate latent (random) pair of (transformed) sensitivity and specificity.  That is
\begin{eqnarray}\label{withinBinom}
Y_{i1}|X_{1}=x_1&\sim& \mbox{Binomial}\Bigl(n_{i1},l^{-1}(x_1)\Bigr);\nonumber\\
Y_{i2}|X_{2}=x_2&\sim& \mbox{Binomial}\Bigl(n_{i2},l^{-1}(x_2)\Bigr),
\end{eqnarray}
where $l(\cdot)$ is a link function.

The stochastic representation of the between studies model takes the form
\begin{equation}\label{copula-between}
\Bigl(F\bigl(X_1;l(\pi_1),\de_1\bigr),F\bigl(X_2;l(\pi_2),\de_2\bigr)\Bigr)\sim C(\cdot;\th),
\end{equation}
where $C(\cdot;\th)$ is a parametric family of copulas with dependence parameter $\th$ and $F(\cdot;l(\pi),\de)$ is the cdf of the univariate distribution of the random effect. The copula parameter $\th$ is a parameter of the random effects model and it is separated from the univariate parameters, the univariate parameters $\pi_1$ and $\pi_2$ are the meta-analytic parameters for the sensitivity and specificity, and  $\de_1$ and $\de_2$  express the variability between studies.
For $N$ studies with data $(y_{ij}, n_{ij}),\, i = 1, \ldots ,N,\, j=1,2$, the models in (\ref{withinBinom}) and (\ref{copula-between}) together specify a copula mixed  model with joint likelihood
\begin{equation}\label{mixed-cop-likelihood}
L(\pi_1,\pi_2,\de_1,\de_2,\th)=\prod_{i=1}^{N}\int_{0}^{1}\int_{0}^{1}
\prod_{j=1}^2g\Bigl(y_{ij};n_{ij},l^{-1}\bigl(F^{-1}(u_j;l(\pi_j),\de_j)\bigr)\Bigr)c(u_1,u_2;\th)du_1du_2,
\end{equation}
where $c(u_1,u_2;\th)=\p^2 C(u_1,u_2;\th)/\p u_1\p u_2$ is the copula  density and $g\bigl(y;n,\pi\bigr)=\binom{n}{y}\pi^y(1-\pi)^{n-y},\quad y=0,1,\ldots,n,\quad 0<\pi<1,$
 is the binomial probability mass function (pmf).
The choices of the  $F\bigl(\cdot;l(\pi),\de\bigr)$ and  $l$ are given in Table \ref{choices}. 

\begin{table}[!h]
\begin{center}
\caption{\label{choices}The choices of the  $F\bigl(\cdot;l(\pi),\de\bigr)$ and  $l$ in the copula mixed model.}
\begin{tabular}{cccc}
\hline $F\bigl(\cdot;l(\pi),\de\bigr)$ & $l$ & $\pi$ & $\de$\\\hline
$N(\mu,\s)$ & logit, probit, cloglog & $l^{-1}(\mu)$&$\s$\\
Beta$(\pi,\gamma)$ & identity & $\pi$ & $\gamma$\\
\hline
\end{tabular}

\end{center}
\end{table}

\subsection{\label{3model}Trivariate copula mixed model}
For each study $i$, the within-study model assumes that the number of true positives $Y_{i1}$, true negatives $Y_{i2}$, and diseased persons $Y_{i3}$ are conditionally independent and binomially distributed given $\X=\x$, where $\X=(X_1,X_2,X_3)$ denotes the  trivariate latent (random) vector of (transformed) sensitivity, specificity, and disease prevalence.  That is
\begin{eqnarray}\label{withinBinom3}
Y_{i1}|X_{1}=x_1&\sim& \mbox{Binomial}\Bigl(n_{i1},l^{-1}(x_1)\Bigr);\nonumber\\
Y_{i2}|X_{2}=x_2&\sim& \mbox{Binomial}\Bigl(n_{i2},l^{-1}(x_2)\Bigr);\\
Y_{i3}|X_{3}=x_3&\sim& \mbox{Binomial}\Bigl(n_{i3},l^{-1}(x_3)\Bigr),\nonumber
\end{eqnarray}
where $l(\cdot)$ is a link function.

The stochastic representation of the between studies model takes the form
\begin{equation}\label{copula-between-norm}
\Bigl(F\bigl(X_1;l(\pi_1),\de_1\bigr),F\bigl(X_2;l(\pi_2),\de_2\bigr)
,F\bigl(X_3;l(\pi_3),\de_3\bigr)\Bigr)\sim C(\cdot;\thbf),
\end{equation}
where $C(\cdot;\thbf)$ is a vine  copula with dependence parameter vector $\thbf=(\th_{12},\th_{13},\th_{23|1})$ and $F(\cdot;l(\pi),\de)$ is the cdf of the  univariate  distribution of the random effect. To be  concrete,  we use the permutation $\{12,13,23|1\}$. The theory though also apply to the other two permutations. 
The joint density $f_{123}(x_1,x_2,x_3)$ of the transformed latent proportions is:
\begin{multline}\label{jointdensityNCMM}
f_{123}(x_1,x_2,x_3;\pi_1,\pi_2,\pi_3,\de_1,\de_2,\de_3,\th,\th_{12},\th_{13})=\\
c_{12}\Bigl(F\bigl(x_1;l(\pi_1),\de_1\bigr),F\bigl(x_2;l(\pi_2),\de_2\bigr)
;\th_{12}\Bigr)\times\\ c_{13}\Bigl(F\bigl(x_1;l(\pi_1),\de_1\bigr),F\bigl(x_3;l(\pi_3),\de_3\bigr);\th_{13}\Bigr)
\prod_{j=1}^3f\bigl(x_j;l(\pi_j),\de_j\bigr),
\end{multline}
where $f(\cdot;l(\pi),\de)$ is the   density of $F$. 

In (\ref{jointdensityNCMM}) we assume conditional independence  between $X_1$ and $X_3$ given $X_2$, i.e., the density of the (independence) copula $C_{13|2}(u,v)=uv$ is $c_{13|2}(u,v)=1$.  Here we are making the simplifying assumption that the conditional copula does not depend on $X_2$. We use simplified vines to keep them tractable for inference and model selection. The simplifying assumption, that copulas of conditional distributions do not depend on the values of the variables which they are conditioned on, is popular \citep{aasetal09} and not restrictive in practice \citep{Stober-joe-czado2013}. 
\cite{joeetal10} show that in order for a (simplified) vine copula to have (tail) dependence for all bivariate margins, it is only necessary the non-conditional  bivariate copulas  to have (tail) dependence and it is not necessary for the conditional bivariate copulas to have tail dependence. That provides the theoretical justification for the idea of conditional independence. For more details see \cite{Nikoloulopoulos2015c}.

For $N$ studies with data $(y_{ij}, n_{ij}),\, i = 1, \ldots ,N,\, j=1,2,3$, the models in (\ref{withinBinom3}) and (\ref{copula-between-norm}) together specify a vine copula mixed  model with joint likelihood
\begin{multline}\label{mixed-cop-likelihood}
L(\pi_1,\pi_2,\pi_3,\de_1,\de_2,\de_3,\th_{12},\th_{13})=\\
\prod_{i=1}^{N}\int_{0}^{1}\int_{0}^{1}\int_{0}^{1}
\prod_{j=1}^3g\Bigl(y_{ij};n_{ij},l^{-1}\bigl(F^{-1}(u_j;l(\pi_j),\de_j)\bigr)\Bigr)c_{12}(u_1,u_2;\th_{12})c_{13}(u_1,u_3;\th_{13})du_j.
\end{multline}
The choices of the  $F\bigl(\cdot;l(\pi),\de\bigr)$ and  $l$ are the same as in  the bivariate case; see Table \ref{choices}.

\subsection{Hybrid copula mixed model}
To form the hybrid copula mixed model we combine the aforementioned models. For ease of exposition, let the first $N_1$ studies be the case-control studies and the remaining $N_2$ studies be the cohort studies. A combination of the  bivariate likelihood for the data from $N_1$ case-control studies and the trivariate likelihood for the data from $N_2$ cohort studies leads to
\begin{multline}\label{beta-mixed-cop-likelihood}
L(\pi_1,\pi_2,\pi_3,\de_1,\de_2,\de_3,\th,\th_{12},\th_{13})=\\
\prod_{i=1}^{N_1}\int_0^1\int_0^1
\prod_{j=1}^2g\Bigl(y_{ij};n_{ij},F^{-1}\bigl(u_j;l(\pi_j),\de_j\bigr)\Bigr)c(u_1,u_2;\th)du_j\quad\times\\
\prod_{i=N_1+1}^{N_1+N_2}\int_0^1\int_0^1\int_0^1
\prod_{j=1}^3g\Bigl(y_{ij};n_{ij},F^{-1}\bigl(u_j;l(\pi_j),\de_j\bigr)\Bigr)c_{12}(u_1,u_2;\th_{12})c_{13}(u_1,u_3;\th_{13})du_j.
\end{multline}

Our general statistical model allows for selection of $c(\cdot;\th)$, $c_{12}(\cdot;\th_{12})$ and $c_{13}(\cdot;\th_{13})$ independently among a variety of parametric copula families, i.e., there are no constraints in the choices of parametric copulas. 

\section{\label{sec-est}Maximum likelihood estimation and computational details}

Estimation of the model parameters $(\pi_1,\pi_2,\pi_3,\de_1,\de_2,\de_3,\th,\th_{12},\th_{13})$    can be approached by the standard ML method, by maximizing the logarithm of the joint likelihood in (\ref{beta-mixed-cop-likelihood}). 
The estimated parameters can be obtained by 
using a quasi-Newton \citep{nash90} method applied to the logarithm of the joint likelihood.  
This numerical  method requires only the objective
function, i.e.,  the logarithm of the joint likelihood, while the gradients
are computed numerically and the Hessian matrix of the second
order derivatives is updated in each iteration. The standard errors (SE) of the ML estimates can be also obtained via the gradients and the Hessian computed numerically during the maximization process.

Numerical evaluation of the mixed joint pmf is easily done with a combination of the algorithms in \cite{Nikoloulopoulos2015b,Nikoloulopoulos2015c}:

\begin{enumerate}
\itemsep=0pt
\item Calculate Gauss-Legendre  quadrature points $\{u_q: q=1,\ldots,n_q\}$ 
and weights $\{w_q: q=1,\ldots,n_q\}$ in terms of standard uniform; see e.g.,  \cite{Stroud&Secrest1966}.
\item 
\begin{enumerate}
\itemsep=0pt
\item Convert from independent uniform random variables $\{u_{q_1}: q_1=1,\ldots,n_q\}$ and $\{u_{q_2}: q_2=1,\ldots,n_q\}$ to dependent uniform random variables $\{u_{q_1}: q_1=1,\ldots,n_q\}$ and $\{C^{-1}(u_{q_2}|u_{q_1};\th): q_1=q_2=1,\ldots,n_q\}$ that have distribution $C(\cdot;\th)$.
The inverse of the conditional distribution $C(v|u;\th)=\partial C(u,v;\th)/\partial u$ corresponding to the copula $C(\cdot;\th)$ is used  to achieve this.

\item Convert from independent uniform random variables $\{u_{q_1}: q_1=1,\ldots,n_q\}$,  $\{u_{q_2}: q_2=1,\ldots,n_q\}$, and $\{u_{q_3}: q_3=1,\ldots,n_q\}$ to dependent uniform random variables $\{v_{q_1}=u_{q_1}: q_1=1,\ldots,n_q\}$, $\bigl\{v_{q_2|q_1}=C^{-1}_{12}(u_{q_2}|u_{q_1};\th_{12}): q_1=q_2=1,\ldots,n_q\bigr\}$, and  
$\Bigl\{v_{q_2q_3|q_1}=C^{-1}_{13}\Bigl(C^{-1}_{23|1}(u_{q_3}|u_{q_2};$ $\th_{23|1}\to 0)|u_{q_1};\th_{13}\Bigr): q_1=q_2=q_3=1,\ldots,n_q\Bigr\}$ that have vine distribution $C(\cdot;\th_{12},\th_{13})$. 
The simulation algorithm of a C-vine copula in \cite{joe2010a} is used  to achieve this.
\end{enumerate}

\item 
\begin{enumerate}
\itemsep=0pt
\item Numerically evaluate the bivariate pmf 
$$\int_0^1\int_0^1
\prod_{j=1}^2g\Bigl(y_{j};n_{j},F^{-1}\bigl(u_j;l(\pi_j),\de_j\bigr)\Bigr)c(u_1,u_2;\th)du_1du_2$$
in a double sum:
$$\sum_{q_1=1}^{n_q}\sum_{q_2=1}^{n_q}w_{q_1}w_{q_2}
g\Bigl(y_1;n_{1},F^{-1}\bigl(u_{q_1};l(\pi_1),\de_1\bigr)\Bigr)g\Bigl(y_{2};n_{2},F^{-1}\bigl(C^{-1}(u_{q_2}|u_{q_1};\th);l(\pi_2),\de_2\bigr)\Bigr).$$

\item Numerically evaluate the trivariate pmf 
$$\int_0^1\int_0^1\int_0^1
\prod_{j=1}^3g\Bigl(y_{ij};n_{ij},F^{-1}\bigl(u_j;l(\pi_j),\de_j\bigr)\Bigr)c_{12}(u_1,u_2;\th_{12})c_{13}(u_1,u_3;\th_{13})du_1du_2du_3$$
in a triple sum 
$$\sum_{q_1=1}^{n_q}\sum_{q_2=1}^{n_q}\sum_{q_3=1}^{n_q}w_{q_1}w_{q_2}w_{q_3}
 \,g\Bigl(y_1;n_{1},F^{-1}\bigl(v_{q_1};l(\pi_1),\de_1\bigr)\Bigr)
 g\Bigl(y_2;n_{2},F^{-1}\bigl(v_{q_2|q_1};l(\pi_2),\de_2\bigr)\Bigr)\times
$$$$g\Bigl(y_3;n_{3},F^{-1}\bigl(v_{q_2q_3|q_1};l(\pi_3),\de_3\bigr)\Bigr)
$$.
\end{enumerate}
\end{enumerate}

\vspace{-1cm}

\noindent The inverse conditional copula cdfs $C^{-1}(v|u;\th)$ are given in Table \ref{2fam}.

With Gauss-Legendre quadrature, the same nodes and weights
are used for different functions;
this helps in yielding smooth numerical derivatives for numerical optimization via quasi-Newton.
Our
extensive comparisons with more quadrature points, show that $n_q=21$ is adequate with good precision to at least at four decimal places.
The developed   algorithm for the calculation of a bivariate or a trivariate integral overcomes the convergence problems that have been reported in the literature \citep{chu-etal-2009,Chen-etal-smmr-2014,chen-etal-2015-jrssc}. Our Gauss-Legendre quadrature algorithm for hybrid copula mixed models (including the hybrid GLMM) is stable. The crucial step is to convert from independent to dependent quadrature points.

\section{\label{sec-sim}Small-sample efficiency--Misspecification}

An extensive simulation study is conducted  
(a) to gauge the small-sample efficiency of the ML 
method, and 
(b) to investigate in detail 
the  misspecification of the parametric margin or  family of copulas of the random effects distributions.

To generate the data we have combined the simulation algorithms in \cite{Nikoloulopoulos2015b,Nikoloulopoulos2015c}:
 
\begin{enumerate}

\item For $i=1,\ldots,N_1$:
\begin{enumerate}

\item Simulate the study size $n$ from a shifted gamma distribution, i.e., $n\sim \mbox{sGamma}(\a=1.2,\b=0.01,\mbox{lag}=30)$ and round off to the nearest integer. 
\item Simulate $(u_1,u_2)$ from a parametric family of copulas $C(;\tau)$;  $\tau$ is converted 
to the copula parameter $\th$ via the relations  in Table \ref{2fam}.  

\item Convert to beta  or normal realizations via $x_j=l^{-1}\Bigl(F_j^{-1}\bigl(u_j,l(\pi_j),\de_j\bigr)\Bigr)$ for $j=1,2$.   
\item Draw the number of diseased $n_{1}$ from a $B(n,0.43)$ distribution.
\item Set  $n_2=n-n_1$, $y_j=n_jx_j$ and then round $y_j$ for $j=1,2$. 
\end{enumerate} 

\item For $i=N_1+1,\ldots,N_1+N_2$
\begin{enumerate}

\item Simulate the study size $n$ from a shifted gamma distribution, i.e., $n\sim \mbox{sGamma}(\a=1.2,\b=0.01,\mbox{lag}=30)$ and round off to the nearest integer. 
\item Simulate $(u_1,u_2,u_3)$ from a C-vine $C(;\tau_{12},\tau_{13},\tau_{23|1}=0)$ via the algorithm  in Joe\cite{joe2010a};  $\tau$'s are converted 
to $\theta$'s via the relations in Table \ref{2fam}. 

\item Convert to beta  or normal realizations via $x_j=l^{-1}\Bigl(F_j^{-1}\bigl(u_j,l(\pi_j),\de_j\bigr)\Bigr)$ for $j=1,2$.    
\item Set number of diseased and non-diseased as $n_{1}=nx_3$ and $n_2=n-n_1$, respectively. 
\item Set   $y_j=n_jx_j$ and then round $y_j$ for $j=1,2$. 
\end{enumerate} 

\end{enumerate}

Tables \ref{sim-norm} and \ref{sim-beta}  contain the 
resultant biases, root mean square errors (RMSE), and standard deviations (SD) for the MLEs  under different copula and marginal choices from $1000$ randomly generated samples of size  $N_1=N_2=25$ from  the hybrid copula mixed model with normal and beta margins, respectively. The true (simulated) copula distributions are the Clayton and Clayton rotated by 90 degrees for the $C_{12}(;\tau_{12})$ and $\{C(;\tau),C_{13}(;\tau_{13})\}$ copulas, respectively.

\setlength{\tabcolsep}{3pt}
\begin{landscape}
\begin{table}[!h]

  \centering
  \caption{\label{sim-norm} Biases,  root mean square errors (RMSE) and standard deviations (SD) for the ML estimates  under different copula choices and margins and CL estimates under normal margins from  small sample of sizes $N_1 = N_2=25$ simulations ($10^3$ replications) from the hybrid copula mixed model with normal margins.
  The true (simulated) copula distributions are the Clayton and Clayton rotated by 90 degrees for the $C_{12}(;\tau_{12})$ and $\{C(;\tau),C_{13}(;\tau_{13})\}$ copulas, respectively.   }
    \begin{tabular}{llccccccccc}
    \hline
   \multicolumn{11}{l}{Biases scaled by 50 for the  estimates  under different copula and margin choices}\\
 \multicolumn{11}{l}{ True  model:  Clayton for $C_{12}(;\tau_{12})$ and Clayton rotated by 90 degrees for $\{C(;\tau),C_{13}(;\tau_{13})\}$ and normal margins} \\
 \multicolumn{2}{l}{True          model        parameters:} &$\pi_1=0.7$ & $\pi_2=0.9$ & $\pi_3=0.7$ & $\s_1=1.5$ & $\s_2=1$ & $\s_3=1.5$ & $\tau_{12}=0.5$ & $\tau_{13}=-0.5$ & $\tau=-0.5$ \\\hline
 \rowcolor{gray}     {Clayton by 0/90} & {Normal} & {0.04} & {-0.39} & {-0.29} & {-1.38} & {-4.12} & {0.01} & {0.83} & {0.92} & {-6.62} \\
     {} & {Beta} & {-2.74} & {-1.96} & {-1.69} & {-} & {-} & {-} & {1.24} & {1.23} & {-5.74} \\
 \rowcolor{gray}     {BVN} & {Normal} & {-0.06} & {-0.39} & {-0.26} & {-2.72} & {-5.46} & {-3.19} & {4.38} & {-2.22} & {-6.47} \\
     {} & {Beta} & {-2.74} & {-1.90} & {-1.63} & {-} & {-} & {-} & {4.76} & {-2.03} & {-5.04} \\
  \rowcolor{gray}    {Clayton by 180/270} & {Normal} & {-0.20} & {-0.32} & {-0.08} & {-2.20} & {-6.17} & {-3.13} & {7.06} & {-2.16} & {-4.97} \\
     {} & {Beta} & {-2.96} & {-1.75} & {-1.22} & {-} & {-} & {-} & {7.38} & {-2.02} & {-2.84} \\
  \rowcolor{gray}   {Independence (CL)} & {Normal} & {-0.64} & {-0.15} & {-0.21} & {-6.08} & {-7.15} & {-5.01} & {-} & {-} & {-} \\\hline
   \multicolumn{11}{l}{SDs scaled by 50 for the  estimates  under different copula and margin choices}\\
 \multicolumn{11}{l}{ True  model:  Clayton for $C_{12}(;\tau_{12})$ and Clayton rotated by 90 degrees for $\{C(;\tau),C_{13}(;\tau_{13})\}$ and normal margins} \\
 \multicolumn{2}{l}{True          model        parameters:} &$\pi_1=0.7$ & $\pi_2=0.9$ & $\pi_3=0.7$ & $\s_1=1.5$ & $\s_2=1$ & $\s_3=1.5$ & $\tau_{12}=0.5$ & $\tau_{13}=-0.5$ & $\tau=-0.5$ \\\hline
  \rowcolor{gray} {Clayton by 0/90} & {Normal} & {1.79} & {0.50} & {3.05} & {8.98} & {6.52} & {11.51} & {12.68} & {7.49} & {6.45} \\
     {} & {Beta} & {1.52} & {0.69} & {2.15} & {2.16} & {1.29} & {2.62 } & {13.85} & {7.14} & {6.32} \\
  \rowcolor{gray} {BVN} & {Normal} & {1.72} & {0.50} & {2.83} & {8.39} & {6.07} & {9.83} & {8.70} & {6.01} & {7.02} \\
     {} & {Beta} & {1.44} & {0.69} & {2.06} & {2.09} & {1.15} & {2.21} & {9.16} & {5.92} & {6.46} \\
 \rowcolor{gray}     {Clayton by 180/270} & {Normal} & {1.75} & {0.51} & {2.86} & {8.67} & {6.39} & {9.62} & {7.10} & {5.20} & {11.47} \\
     {} & {Beta} & {1.47} & {0.69} & {2.13} & {2.16} & {1.16} & {2.16} & {6.73} & {5.35} & {10.49} \\
 \rowcolor{gray}     {Independence (CL)} & {Normal} & {2.22} & {0.63} & {3.41} & {7.77} & {5.96} & {9.67} & {-} & {-} & {-} \\
 \hline
   \multicolumn{11}{l}{RMSEs scaled by 50 for the  estimates  under different copula and margin choices}\\
 \multicolumn{11}{l}{ True  model:  Clayton for $C_{12}(;\tau_{12})$ and Clayton rotated by 90 degrees for $\{C(;\tau),C_{13}(;\tau_{13})\}$ and normal margins} \\
 \multicolumn{2}{l}{True          model        parameters:} &$\pi_1=0.7$ & $\pi_2=0.9$ & $\pi_3=0.7$ & $\s_1=1.5$ & $\s_2=1$ & $\s_3=1.5$ & $\tau_{12}=0.5$ & $\tau_{13}=-0.5$ & $\tau=-0.5$ \\\hline
 \rowcolor{gray}     {Clayton by 0/90} & {Normal} & {1.79} & {0.64} & {3.07} & {9.08} & {7.71} & {11.51} & {12.71} & {7.55} & {9.24} \\
    {} & {Beta} & {3.13} & {2.08} & {2.74} & {-} & {-} & {-} & {13.90} & {7.25} & {8.54} \\
   \rowcolor{gray}   {BVN} & {Normal} & {1.72} & {0.64} & {2.84} & {8.82} & {8.16} & {10.34} & {9.74} & {6.40} & {9.54} \\
    {} & {Beta} & {3.10} & {2.02} & {2.63} & {-} & {-} & {-} & {10.32} & {6.25} & {8.19} \\
\rowcolor{gray}   {Clayton by 180/270} & {Normal} & {1.76} & {0.60} & {2.86} & {8.95} & {8.88} & {10.12} & {10.02} & {5.63} & {12.50} \\
    {} & {Beta} & {3.31} & {1.88} & {2.45} & {-} & {-} & {-} & {9.99} & {5.72} & {10.87} \\
   \rowcolor{gray}   {Independence (CL)} & {Normal} & {2.31} & {0.65} & {3.42} & {9.87} & {9.31} & {10.89} & {-} & {-} & {-} \\
   \hline
    \end{tabular}
   
    \begin{footnotesize}
Clayton by $\alpha/\beta$ denotes a hybrid copula mixed with copula distributions  the Clayton rotated by $\alpha$ and $\beta$ degrees for the $C_{12}(;\tau_{12})$ and $\{C(;\tau),C_{13}(;\tau_{13})\}$ copulas, respectively.
\end{footnotesize}
\end{table}
\end{landscape}

\setlength{\tabcolsep}{3pt}
\begin{landscape}
\begin{table}[!h]
  \centering
  \caption{\label{sim-beta} Biases,  root mean square errors (RMSE) and standard deviations (SD) for the ML estimates  under different copula choices and margins and CL estimates under beta margins from  small sample of sizes $N_1 = N_2=25$ simulations ($10^3$ replications) from the hybrid copula mixed model with normal margins.
  The true (simulated) copula distributions are the Clayton and Clayton rotated by 90 degrees for the $C_{12}(;\tau_{12})$ and $\{C(;\tau),C_{13}(;\tau_{13})\}$ copulas, respectively.  }
    \begin{tabular}{llccccccccc}
\hline
   \multicolumn{11}{l}{Biases scaled by 50 for the  estimates  under different copula and margin choices}\\
 \multicolumn{11}{l}{ True  model:  Clayton for $C_{12}(;\tau_{12})$ and Clayton rotated by 90 degrees for $\{C(;\tau),C_{13}(;\tau_{13})\}$ and beta margins} \\
 \multicolumn{2}{l}{True          model        parameters:} &$\pi_1=0.7$ & $\pi_2=0.9$ & $\pi_3=0.7$ & $\g_1=0.15$ & $\g_2=0.1$ & $\g_3=0.15$ & $\tau_{12}=0.5$ & $\tau_{13}=-0.5$ & $\tau=-0.5$ \\\hline    
    Clayton by 0/90 & Normal & 1.64  & 1.43  & 0.66  & -     & -     & -     & 2.17  & 0.45  & -7.40 \\
 \rowcolor{gray}                & Beta  & 0.00  & -0.16 & -0.19 & -0.39 & -0.58 & -0.09 & 2.38  & 0.90  & -7.43 \\
         BVN   & Normal & 1.57  & 1.46  & 0.76  & -     & -     & -     & 4.95  & -2.49 & -7.04 \\
   \rowcolor{gray}          & Beta  & 0.02  & -0.09 & -0.09 & -0.75 & -0.85 & -0.72 & 4.98  & -2.53 & -6.49 \\
          Clayton by 180/270 & Normal & 1.45  & 1.47  & 0.82  & -     & -     & -     & 7.12  & -3.54 & -6.36 \\
\rowcolor{gray}               & Beta  & -0.08 & -0.05 & 0.09  & -0.62 & -0.89 & -0.69 & 7.36  & -3.61 & -4.93 \\
           Independence (CL)    & Normal & 1.38  & 1.59  & 0.80  & -     & -     & -     & -     & -     &  \\
 \hline
   \multicolumn{11}{l}{SDs scaled by 50 for the  estimates  under different copula and margin choices}\\
 \multicolumn{11}{l}{ True  model:  Clayton for $C_{12}(;\tau_{12})$ and Clayton rotated by 90 degrees for $\{C(;\tau),C_{13}(;\tau_{13})\}$ and beta margins} \\
 \multicolumn{2}{l}{True          model        parameters:} &$\pi_1=0.7$ & $\pi_2=0.9$ & $\pi_3=0.7$ & $\g_1=0.15$ & $\g_2=0.1$ & $\g_3=0.15$ & $\tau_{12}=0.5$ & $\tau_{13}=-0.5$ & $\tau=-0.5$ \\\hline   
    Clayton by 0/90 & Normal & 1.08  & 0.46  & 1.90  & 7.05  & 8.17  & 8.23  & 11.66 & 9.29  & 6.71 \\
 \rowcolor{gray}               & Beta  & 1.04  & 0.57  & 1.64  & 1.56  & 1.19  & 2.01  & 12.89 & 9.21  & 6.62 \\
           BVN   & Normal & 1.03  & 0.45  & 1.76  & 6.13  & 7.49  & 6.94  & 8.03  & 6.61  & 7.67 \\
 \rowcolor{gray}                & Beta  & 0.99  & 0.54  & 1.53  & 1.35  & 1.05  & 1.58  & 8.38  & 6.51  & 7.38 \\
           Clayton by 180/270 & Normal & 1.09  & 0.46  & 1.76  & 6.02  & 7.88  & 7.00  & 7.49  & 5.76  & 12.37 \\
 \rowcolor{gray}                & Beta  & 1.05  & 0.54  & 1.56  & 1.42  & 1.10  & 1.55  & 7.12  & 5.87  & 12.10 \\
           Independence (CL)    & Normal & 1.41  & 0.57  & 2.13  & 5.64  & 7.33  & 6.70  &       &       &  \\
\hline
   \multicolumn{11}{l}{RMSEs scaled by 50 for the  estimates  under different copula and margin choices}\\
 \multicolumn{11}{l}{ True  model:  Clayton for $C_{12}(;\tau_{12})$ and Clayton rotated by 90 degrees for $\{C(;\tau),C_{13}(;\tau_{13})\}$ and beta margins} \\
 \multicolumn{2}{l}{True          model        parameters:} &$\pi_1=0.7$ & $\pi_2=0.9$ & $\pi_3=0.7$ & $\g_1=0.15$ & $\g_2=0.1$ & $\g_3=0.15$ & $\tau_{12}=0.5$ & $\tau_{13}=-0.5$ & $\tau=-0.5$ \\\hline   
     Clayton by 0/90 & Normal & 1.96  & 1.50  & 2.01  & -     & -     & -     & 11.86 & 9.30  & 9.99 \\
    \rowcolor{gray}             & Beta  & 1.04  & 0.59  & 1.65  & 1.61  & 1.33  & 2.01  & 13.11 & 9.26  & 9.95 \\
          BVN   & Normal & 1.88  & 1.53  & 1.92  & -     & -     & -     & 9.43  & 7.06  & 10.41 \\
  \rowcolor{gray}               & Beta  & 0.99  & 0.55  & 1.53  & 1.55  & 1.35  & 1.74  & 9.75  & 6.98  & 9.83 \\
           Clayton by 180/270& Normal & 1.81  & 1.54  & 1.94  & -     & -     & -     & 10.33 & 6.76  & 13.91 \\
    \rowcolor{gray}            & Beta  & 1.06  & 0.55  & 1.56  & 1.55  & 1.41  & 1.70  & 10.24 & 6.89  & 13.06 \\
         Independence (CL)   & Normal & 1.97  & 1.69  & 2.28  & -     & -     & -     & -     & -     & - \\
    \hline
    \end{tabular}

\begin{footnotesize}
Clayton by $\alpha/\beta$ denotes a hybrid copula mixed with copula distributions  the Clayton rotated by $\alpha$ and $\beta$ degrees for the $C_{12}(;\tau_{12})$ and $\{C(;\tau),C_{13}(;\tau_{13})\}$ copulas, respectively.
\end{footnotesize}
\end{table}
\end{landscape}

We also report these summaries for the CL estimates in \cite{chen-etal-2015-jrssc} to allow for a comprehensive comparison.  
In \cite{chen-etal-2015-jrssc} it has been  assumed that the association $\tau_{12}$ between sensitivity and specificity for cohort studies is the same as the association   $\tau$ between sensitivity and specificity for case-control studies, i.e., $\tau=\tau_{12}$. This is a  strong assumption given the fact that the sensitivity/specificity depends on disease prevalence in cohort studies, thus the association between sensitivity and specificity is likely to be affected.   In our simulations   we emphasize  that by allowing heterogeneity in association in  cohort and case control studies. 
Any comparison of the likelihood methods in terms of computing time is a digression and not included here. It is obvious that the CL method is faster than the ML method and not in need of a comparison, since the idea is to replace a numerically more difficult high-dimensional probability calculation with a much simpler probability calculation assuming independence among random effects.

Conclusions from the values in the tables are the following:

\begin{itemize}
\item ML   with  the true hybrid copula mixed  model is highly efficient according to the simulated biases and standard deviations.

\item The CL method yields estimates that are almost as good as the ML estimates for the meta-analytic parameters under the assumption of normal margins. 

\item The CL method slightly underestimates the between-studies variability parameters.

\item The ML estimates of the meta-analytic parameters are slightly underestimated under copula misspecification.

\item The SDs are rather robust to the copula misspecification.

\item The meta-analytic ML and CL estimates are not robust to the margin misspecification, while 
the ML estimate of $\tau$  is.

\end{itemize}

The meta-analytic parameters are a univariate inference, and hence it is the univariate marginal distribution that  matters and not the type of the copula; see also  \cite{Nikoloulopoulos2015b,
Nikoloulopoulos2015c}.  \cite{chen-etal-2015-jrssc} constraint   themselves to normal margins; this it is too restrictive and as shown in Table \ref{sim-beta} leads to overestimation of the meta-analytic parameters when the true univariate distribution  of the latent sensitivity, specificity, and disease prevalence  is  beta.

\section{\label{sec-app}Systematic review of modern diagnostic imaging modalities for surveillance of melanoma patients}
To assess the diagnostic imaging modalities for the surveillance of melanoma patients we apply hybrid copula mixed models. 
The diagnostic modalities under investigation are ultrasonography (US) for regional lymph node metastasis ($N_1=6,N_2=15$) and  
positron emission tomography (PET) for both  regional ($N_1=5,N_2=17$)   and distant ($N_1=15,N_2=15$)  lymph node metastasis. 
We fit the hybrid copula mixed model for all different permutations, choices of parametric families of copulas and margins.
To make it easier
to compare strengths of dependence, we convert from $\theta$'s to $\tau$'s via the relations in Table \ref{2fam}. 
Since the number of parameters is the same between the models, we  use the maximized  log-likelihood that corresponds to the  estimates as a rough diagnostic measure for goodness of fit between the  models. We also estimate the model parameters with the CL method in \cite{chen-etal-2015-jrssc}.
Finally, we  demonstrate summary receiver operating characteristic  (SROC)  curves and summary operating points (a pair of average sensitivity and specificity) with a confidence region and a predictive region  \citep{Nikoloulopoulos2015b}.

\begin{figure}[!h]
\begin{center}
\begin{footnotesize}
\begin{tabular}{|cc|}
\hline

Cohort studies& Case-Control studies\\\hline
\includegraphics[width=0.5\textwidth]{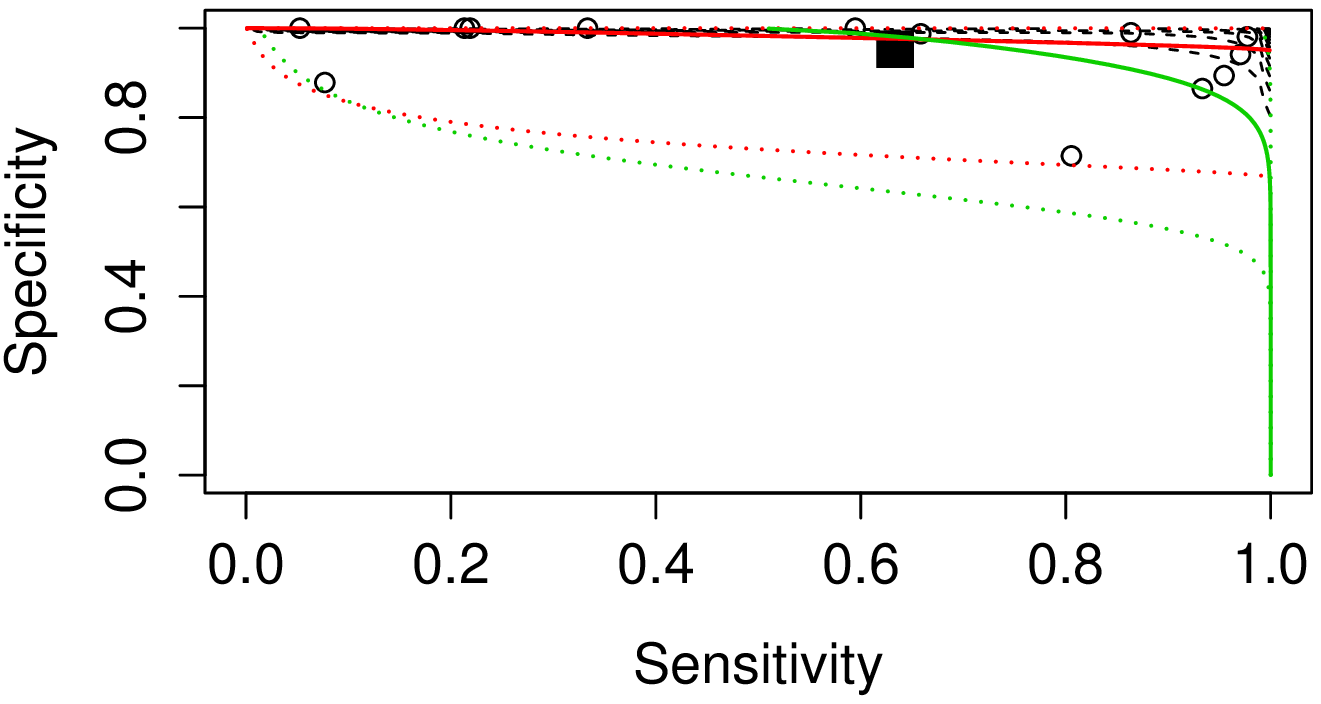}
&

\includegraphics[width=0.5\textwidth]{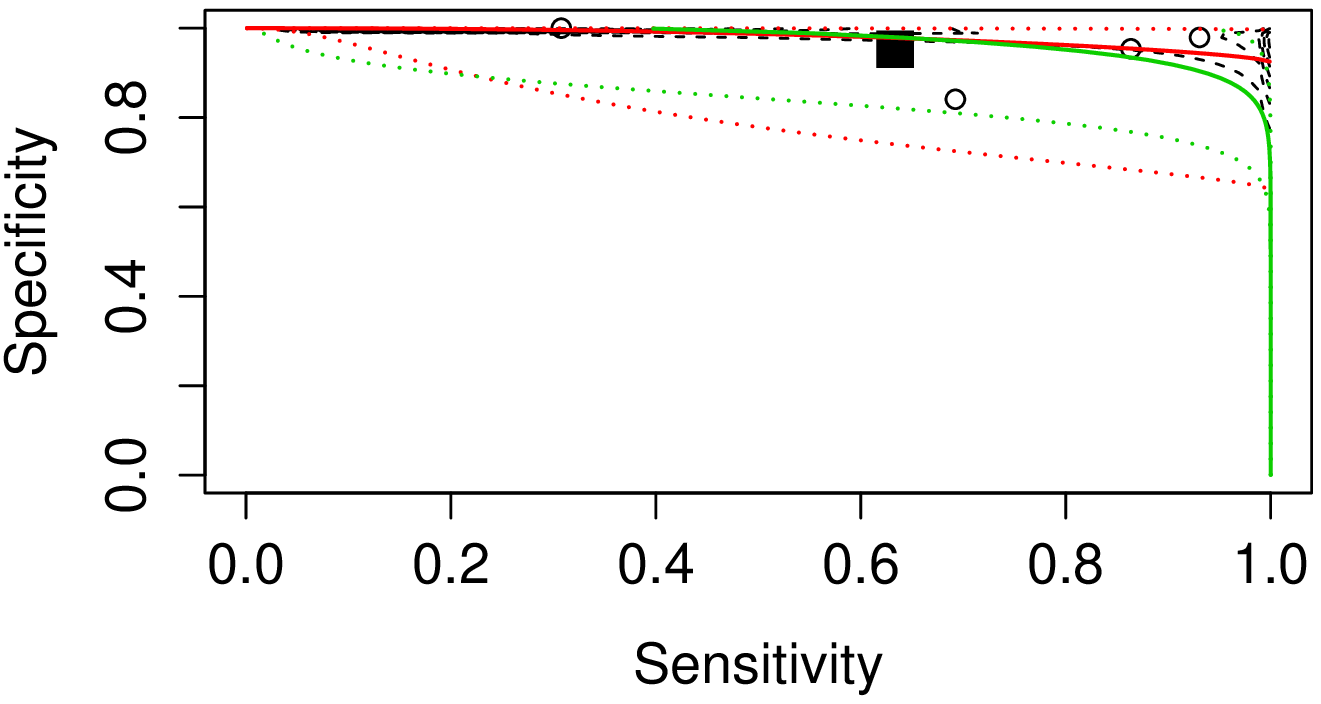}\\
\hline\end{tabular}
\end{footnotesize}
\caption{\label{US-SROC}Contour plots (predictive region)  and quantile  regression curves  from the hybrid copula mixed model with copula distributions  the Clayton rotated by 180 and 270 degrees for the $C_{13}(;\tau_{13})$ and $\{C(;\tau),C_{13}(;\tau_{13})\}$ copulas, respectively and    beta margins for the US modality to diagnose regional lymph node metastasis. Red and green lines  represent the quantile  regression curves $x_1:=\widetilde{x}_1(x_2,q)$ and $x_2:=\widetilde{x}_2(x_1,q)$, respectively; for $q=0.5$ solid lines and for $q\in\{0.01,0.99\}$ dotted lines (confidence region).}
\end{center}
\vspace{-0.5cm}
\end{figure}

In Table \ref{US} we report the resulting maximized log-likelihoods, estimates, and standard errors of the hybrid copula mixed models with different choices of parametric families of copulas and margins for the US modality to diagnose regional lymph node metastasis. 
All models roughly agree on the estimated sensitivity $\hat\pi_1$ and specificity $\hat\pi_2$,  but the estimate of disease prevalence  is higher when beta margins are assumed. In fact, the log-likelihoods show that a hybrid copula mixed with copula distributions  the Clayton rotated by 180 and 270 degrees for the $C_{13}(;\tau_{13})$ and $\{C(;\tau),C_{12}(;\tau_{12})\}$ copulas, respectively, and    beta margins provides the best fit. 
It is also provides better inferences than a hybrid copula-based mixed model with independence among the random effects since the likelihood has been improved by $6.9=-190.98-(-197.88)$ units. This is also confirmed by a likelihood ratio test ($p$-value $\leq 0.001$). 
Hence apparently, the CL method in Chen et al.\cite{chen-etal-2015-jrssc} underestimates the disease prevalence of metastases. This has also to do with the incorrect assumption  of a normal margin in addition to the assumption of independence among random effects. Figure \ref{US-SROC}  shows  the fitted SROC curves along with their confidence and prediction regions for  the best fitted hybrid copula mixed model with beta margins for both case-control and cohort studies.

\setlength{\tabcolsep}{4pt}
\begin{table}[!h]
  \centering
  \caption{\label{US}Maximised  ML and CL log-likelihoods, estimates and standard errors (SE)  of the hybrid copula mixed models with different choices of parametric families of copulas and margins for the US modality to diagnose regional lymph node metastasis.}
  \vspace{1cm}
    \begin{tabular}{ccccccccccccccc}
    \hline
          & \multicolumn{14}{c}{Normal margins} \\
    \hline
          & \multicolumn{2}{c}{BVN} & \multicolumn{2}{c}{Frank} & \multicolumn{2}{c}{Cln 180/270} & \multicolumn{2}{c}{Cln 180/90} & \multicolumn{2}{c}{Cln 0/270} & \multicolumn{2}{c}{Cln 0/90} & \multicolumn{2}{c}{CL} \\
          & Est.  & SE    & Est.  & SE    & Est.  & SE    & Est.  & SE    & Est.  & SE    & Est.  & SE    & Est.  & SE \\\hline
    $\pi_1$ & 0.68  & 0.11  & 0.69  & 0.10  & 0.71  & 0.10  & 0.68  & 0.11  & 0.74  & 0.09  & 0.64  & 0.13  & 0.68  & 0.11 \\
    $\pi_2$ & 0.98  & 0.01  & 0.98  & 0.01  & 0.98  & 0.01  & 0.98  & 0.01  & 0.98  & 0.01  & 0.98  & 0.01  & 0.98  & 0.01 \\
    $\pi_3$ & 0.19  & 0.06  & 0.15  & 0.06  & 0.15  & 0.05  & 0.15  & 0.05  & 0.13  & 0.05  & 0.18  & 0.09  & 0.15  & 0.08 \\
    $\s_1$ & 1.94  & 0.37  & 1.91  & 0.37  & 1.97  & 0.42  & 2.01  & 0.43  & 1.92  & 0.41  & 2.03  & 0.43  & 1.97  & 0.14 \\
    $\s_2$ & 1.54  & 0.37  & 1.54  & 0.36  & 1.53  & 0.35  & 1.56  & 0.41  & 1.53  & 0.35  & 1.53  & 0.38  & 1.42  & 0.06 \\
    $\s_3$ & 2.53  & 0.57  & 2.59  & 0.57  & 2.56  & 0.55  & 2.56  & 0.55  & 2.57  & 0.56  & 2.54  & 0.57  & 2.58  & 0.15 \\
    $\tau_{12}$ & -0.26 & 0.16  & -0.31 & 0.17  & -0.28 & 0.17  & -0.21 & 0.19  & -0.28 & 0.16  & -0.21 & 0.16  & 0.00  & - \\
    $\tau_{13}$ & 0.01  & 0.16  & -0.05 & 0.20  & -0.15 & 0.17  & -0.13 & 0.24  & -0.25 & 0.11  & 0.09  & 0.16  & 0.00  & - \\
    $\tau$ & -0.40 & 0.34  & -0.37 & 0.35  & -0.47 & 0.30  & -0.22 & 0.46  & -0.46 & 0.31  & -0.23 & 0.43  & 0.00  & - \\\hline
    $\log L$ & \multicolumn{2}{c}{-194.38} & \multicolumn{2}{c}{193.94} & \multicolumn{2}{c}{-193.59} & \multicolumn{2}{c}{-194.96} & \multicolumn{2}{c}{-193.05} & \multicolumn{2}{c}{-195.22} & \multicolumn{2}{c}{-197.88} \\
    \hline
\end{tabular}  

\begin{center}
 \begin{tabular}{ccccccccccccc}
    
          & \multicolumn{12}{c}{Beta margins} \\
    \hline
           & \multicolumn{2}{c}{BVN} & \multicolumn{2}{c}{Frank} & \multicolumn{2}{c}{Cln 180/270} & \multicolumn{2}{c}{Cln 180/90} & \multicolumn{2}{c}{Cln 0/270} & \multicolumn{2}{c}{Cln 0/90}  \\
          & Est.  & SE    & Est.  & SE    & Est.  & SE    & Est.  & SE    & Est.  & SE    & Est.  & SE \\\hline
    $\pi_1$ & 0.61  & 0.07  & 0.61  & 0.07  & 0.63  & 0.07  & 0.61  & 0.07  & 0.63  & 0.07  & 0.60  & 0.08 \\
    $\pi_2$ & 0.95  & 0.02  & 0.96  & 0.01  & 0.95  & 0.01  & 0.95  & 0.02  & 0.95  & 0.01  & 0.95  & 0.02 \\
    $\pi_3$ & 0.28  & 0.05  & 0.27  & 0.06  & 0.26  & 0.07  & 0.26  & 0.05  & 0.29  & 0.06  & 0.27  & 0.06 \\
    $\g_1$ & 0.36  & 0.07  & 0.36  & 0.07  & 0.38  & 0.08  & 0.38  & 0.08  & 0.36  & 0.07  & 0.37  & 0.08 \\
    $\g_2$ & 0.10  & 0.04  & 0.09  & 0.04  & 0.09  & 0.04  & 0.10  & 0.05  & 0.09  & 0.04  & 0.10  & 0.05 \\
    $\g_3$ & 0.35  & 0.09  & 0.36  & 0.08  & 0.35  & 0.08  & 0.35  & 0.08  & 0.35  & 0.08  & 0.36  & 0.09 \\
    $\tau_{12}$ & -0.27 & 0.16  & -0.32 & 0.16  & -0.30 & 0.16  & -0.19 & 0.16  & -0.30 & 0.16  & -0.19 & 0.17 \\
    $\tau_{13}$ & -0.04 & 0.18  & -0.01 & 0.14  & -0.19 & 0.12  & -0.16 & 0.14  & -0.08 & 0.15  & 0.00  & 0.16 \\
    $\tau$ & -0.42 & 0.33  & -0.38 & 0.34  & -0.49 & 0.29  & -0.25 & 0.43  & -0.48 & 0.29  & -0.23 & 0.44 \\\hline
    $\log L$ & \multicolumn{2}{c}{-191.98} & \multicolumn{2}{c}{-191.68} & \multicolumn{2}{c}{-190.98} & \multicolumn{2}{c}{-192.68} & \multicolumn{2}{c}{-191.24} & \multicolumn{2}{c}{-192.95} \\
    \hline
    \end{tabular}
\end{center}
\begin{footnotesize}
Cln $\alpha/\beta$ denotes a hybrid copula mixed with copula distributions  the Clayton rotated by $\alpha$ and $\beta$ degrees for the $C_{13}(;\tau_{13})$ and $\{C(;\tau),C_{12}(;\tau_{12})\}$ copulas, respectively.
\end{footnotesize}
\end{table}

\begin{figure}[!h]
\begin{center}
\begin{footnotesize}
\begin{tabular}{|cc|}
\hline

Cohort studies& Case-Control studies\\\hline
\includegraphics[width=0.5\textwidth]{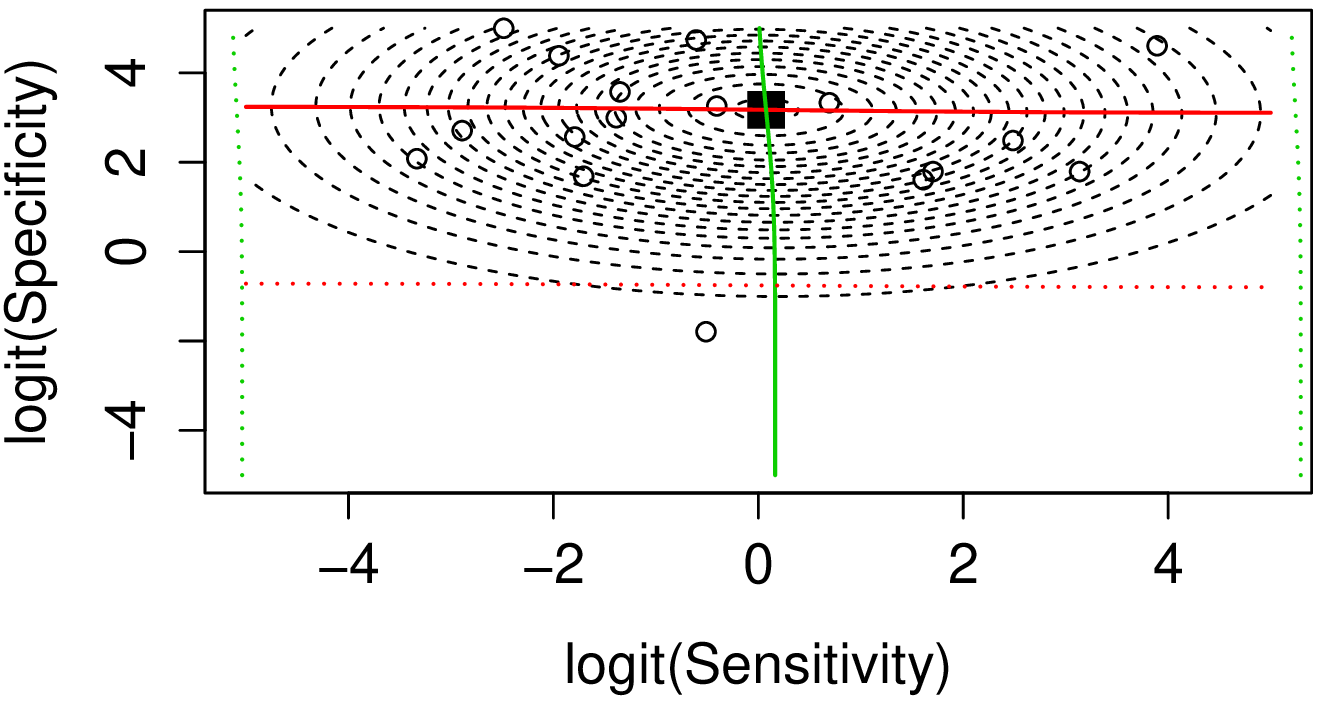}
&

\includegraphics[width=0.5\textwidth]{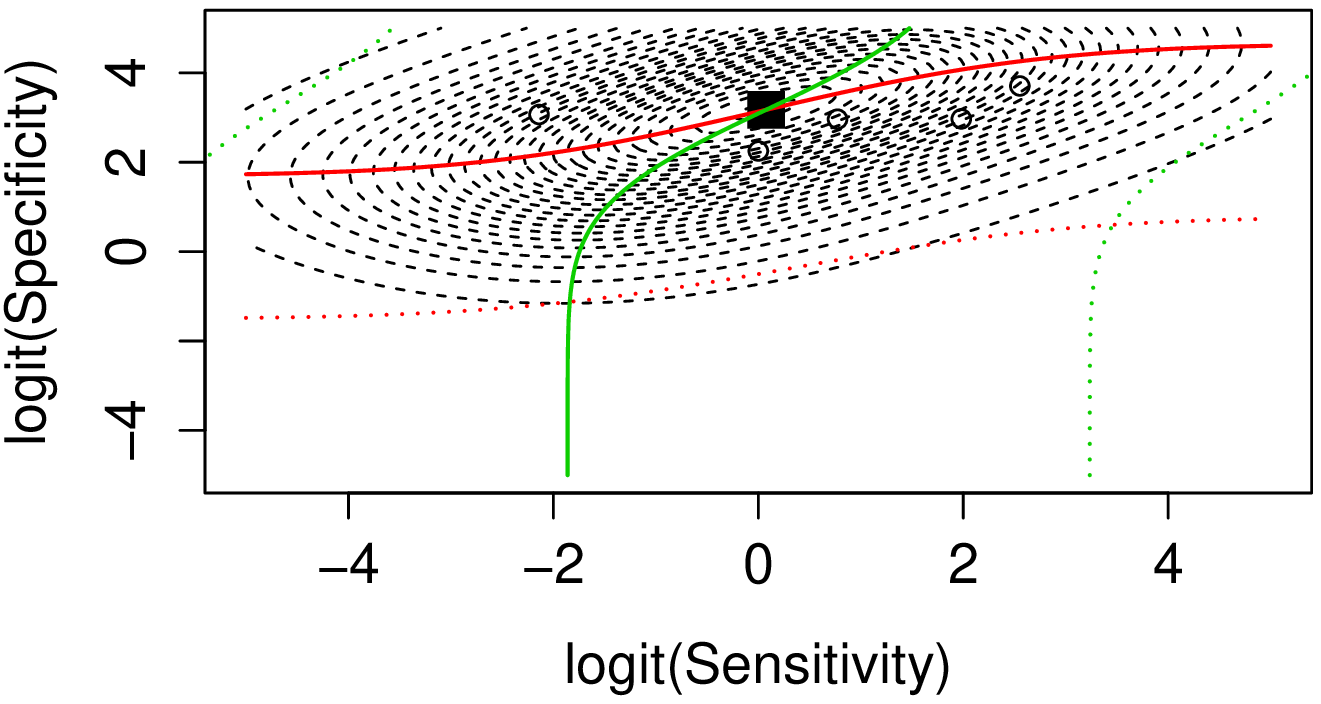}\\
\hline\end{tabular}
\end{footnotesize}
\caption{\label{PET-regional-SROC}Contour plots (predictive region)  and quantile  regression curves  from the hybrid copula mixed model with Frank copulas  and    normal margins for the PET modality to diagnose regional lymph node metastasis. Red and green lines represent the quantile  regression curves $x_1:=\widetilde{x}_1(x_2,q)$ and $x_2:=\widetilde{x}_2(x_1,q)$, respectively; for $q=0.5$ solid lines and for $q\in\{0.01,0.99\}$ dotted lines (confidence region).}
\end{center}
\vspace{-0.5cm}
\end{figure}

In Table \ref{PET-CT-regional} we report the resulting maximized log-likelihoods, estimates, and standard errors of the hybrid copula mixed models with different choices of parametric families of copulas and margins for the PET 
modality to diagnose regional lymph node metastasis.  
All models roughly agree on the estimated sensitivity $\hat\pi_1$,  and disease prevalence $\hat\pi_3$, but the estimate $\hat\pi_2$  of specificity is smaller when beta margins are assumed. The log-likelihoods show that a hybrid copula mixed model with Frank copulas and normal margins provides the best fit.
It is also provides better inferences than a hybrid copula-based mixed model with independence among the random effects since the likelihood has been improved by $3.86=-154.12-(-157.98)$ units. This is also confirmed by a likelihood ratio test ($p$-value $=0.005$). 
Figure \ref{PET-regional-SROC}  shows  the fitted SROC curves along with their confidence and prediction regions for  the best fitted hybrid copula mixed model with normal margins for both case-control and cohort studies.

\setlength{\tabcolsep}{8pt}
\begin{table}[!h]
  \centering
  \caption{\label{PET-CT-regional}Maximised  ML and CL log-likelihoods, estimates and standard errors (SE)  of the hybrid copula mixed models with different choices of parametric families of copulas and margins for the PET 
  modality to diagnose 
  regional lymph node metastasis.}
    \begin{tabular}{ccccccccccc}
    \hline
          & \multicolumn{10}{c}{Normal margins} \\
    \hline
          & \multicolumn{2}{c}{BVN} & \multicolumn{2}{c}{Frank} & \multicolumn{2}{c}{Clayton} & \multicolumn{2}{c}{Clayton  180} & \multicolumn{2}{c}{CL} \\
          & Est.  & SE    & Est.  & SE    & Est.  & SE    & Est.  & SE    & Est.  & SE \\\hline
    {$\pi_1$} & 0.50  & 0.14  & 0.52  & 0.14  & 0.50  & 0.13  & 0.56  & 0.14  & 0.47  & 0.13 \\
    {$\pi_2$} & 0.96  & 0.02  & 0.96  & 0.02  & 0.96  & 0.02  & 0.95  & 0.02  & 0.97  & 0.02 \\
    {$\pi_3$} & 0.37  & 0.04  & 0.37  & 0.04  & 0.36  & 0.04  & 0.37  & 0.04  & 0.35  & 0.05 \\
    {$\s_1$} & 2.33  & 0.55  & 2.22  & 0.53  & 2.32  & 0.50  & 2.44  & 0.53  & 2.27  & 0.13 \\
    {$\s_2$} & 1.67  & 0.54  & 1.69  & 0.58  & 1.72  & 0.52  & 1.51  & 0.45  & 1.75  & 0.29 \\
    {$\s_3$} & 0.65  & 0.16  & 0.64  & 0.15  & 0.68  & 0.16  & 0.64  & 0.16  & 0.71  & 0.04 \\
    {$\tau_{12}$} & 0.02  & 0.18  & -0.01 & 0.20  & 0.00  & 0.12  & -0.05 & 0.19  & 0.00  & - \\
    {$\tau_{13}$} & 0.54  & 0.17  & 0.54  & 0.16  & 0.60  & 0.17  & 0.55  & 0.20  & 0.00  & - \\
    {$\tau$} & 0.50  & 0.45  & 0.35  & 0.79  & 0.38  & 0.43  & 0.63  & 0.36  & 0.00  & - \\\hline
    {$\log L$} & \multicolumn{2}{c}{-154.45} & \multicolumn{2}{c}{-154.12} & \multicolumn{2}{c}{-154.80} & \multicolumn{2}{c}{-154.33} & \multicolumn{2}{c}{-157.98} \\
    \hline
    \end{tabular}
    \begin{tabular}{ccccccccc}
    
          & \multicolumn{8}{c}{Beta margins} \\
    \hline
          & \multicolumn{2}{c}{BVN} & \multicolumn{2}{c}{Frank} & \multicolumn{2}{c}{Clayton} & \multicolumn{2}{c}{Clayton  180} \\
          & Est.  & SE    & Est.  & SE    & Est.  & SE    & Est.  & SE \\\hline
    {$\pi_1$} & 0.50  & 0.07  & 0.51  & 0.07  & 0.49  & 0.07  & 0.49  & 0.07 \\
    {$\pi_2$} & 0.89  & 0.03  & 0.89  & 0.04  & 0.89  & 0.03  & 0.90  & 0.03 \\
    {$\pi_3$} & 0.38  & 0.04  & 0.38  & 0.04  & 0.38  & 0.04  & 0.37  & 0.04 \\
    {$\g_1$} & 0.44  & 0.09  & 0.42  & 0.08  & 0.44  & 0.08  & 0.43  & 0.07 \\
    {$\g_2$} & 0.22  & 0.07  & 0.22  & 0.07  & 0.22  & 0.07  & 0.24  & 0.08 \\
    {$\g_3$} & 0.08  & 0.04  & 0.08  & 0.03  & 0.09  & 0.04  & 0.07  & 0.03 \\
    {$\tau_{12}$} & 0.02  & 0.18  & -0.02 & 0.20  & 0.00  & 0.10  & 0.04  & 0.20 \\
    {$\tau_{13}$} & 0.55  & 0.17  & 0.54  & 0.16  & 0.60  & 0.17  & 0.52  & 0.19 \\
    {$\tau$} & 0.44  & 0.44  & 0.26  & 0.77  & 0.36  & 0.39  & -0.32 & 0.14 \\\hline
    {$\log L$} & \multicolumn{2}{c}{-157.36} & \multicolumn{2}{c}{-157.03} & \multicolumn{2}{c}{-157.69} & \multicolumn{2}{c}{-156.74} \\
    \hline
    \end{tabular}
  \label{tab:addlabel}
\end{table}

Comparing the results in Tables \ref{US} and \ref{PET-CT-regional}   for the surveillance of regional lymph node metastasis, US has the highest sensitivity (63\%; 95\% confidence interval CI = 50--77\%) and specificity (95\%; 95\% CI = 93--97\%). In contrast, patients diagnosed by PET  have higher estimated prevalences of metastasis (37\%; 95\% confidence interval CI = 29--45\%), compared with patients diagnosed by US (26\%; 95\% confidence interval CI = 12--40\%).

\begin{figure}[!h]
\begin{center}
\begin{footnotesize}
\begin{tabular}{|cc|}
\hline

Cohort studies& Case-Control studies\\\hline
\includegraphics[width=0.5\textwidth]{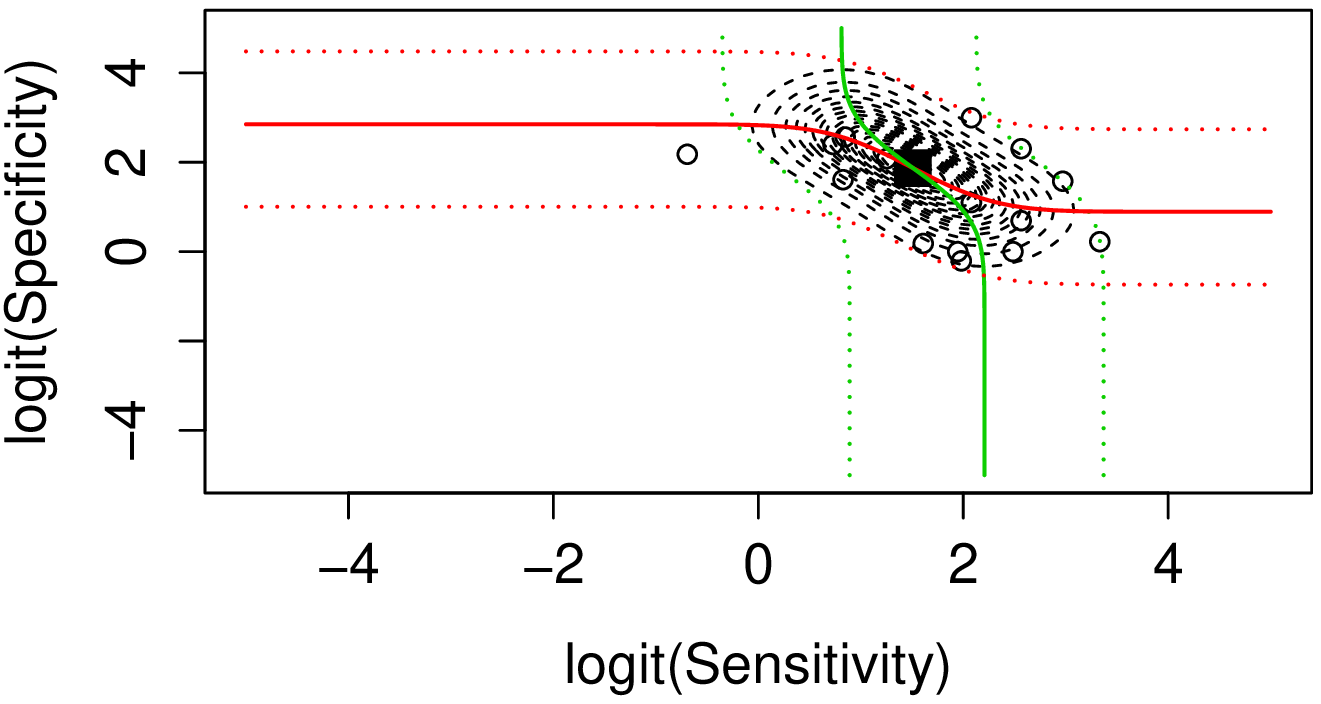}
&

\includegraphics[width=0.5\textwidth]{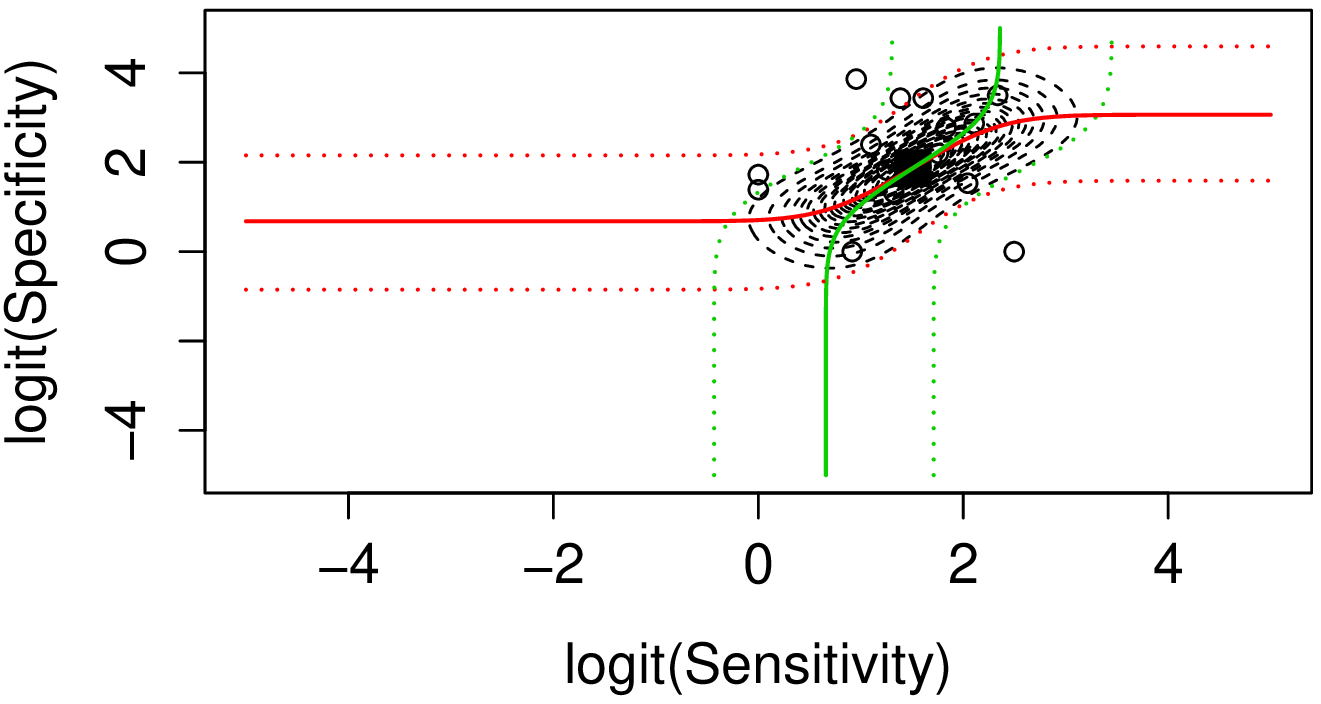}\\
\hline\end{tabular}
\end{footnotesize}
\caption{\label{PET-distant-SROC}Contour plots (predictive region)  and quantile  regression curves  from the hybrid copula mixed model with Frank copulas and    normal margins for the PET  modality to diagnose distant lymph node metastasis. Red and green lines represent the quantile  regression curves $x_1:=\widetilde{x}_1(x_2,q)$ and $x_2:=\widetilde{x}_2(x_1,q)$, respectively; for $q=0.5$ solid lines and for $q\in\{0.01,0.99\}$ dotted lines (confidence region).}
\end{center}
\vspace{-0.5cm}
\end{figure}

Finally, in  Table \ref{PET-CT-distant} we report the resulting maximized log-likelihoods, estimates, and standard errors of the hybrid copula mixed models with different choices of parametric families of copulas and margins for the for the PET  
 modality to diagnose 
 distant 
 lymph node metastasis. 
All models roughly agree on the estimated sensitivity $\hat\pi_1$, specificity  $\hat\pi_2$  and disease prevalence $\hat\pi_3$ for the surveillance of regional lymph node metastasis. The log-likelihoods show that a hybrid copula mixed model with Frank copulas and normal margins provides the best fit. 
It is also provides better inferences than a hybrid copula-based mixed model with independence among the random effects since the likelihood has been improved by $4.34=-174.46-(-178.8)$ units. This is also confirmed by a likelihood ratio test ($p$-value $=0.003$). 
Figure \ref{PET-distant-SROC}  shows  the fitted SROC curves along with their confidence and prediction regions for  the best fitted hybrid copula mixed model with normal margins for both case-control and cohort studies. 
In this dataset it revealed  that there is heterogeneity in association between cohort and case control studies, i.e. $\hat\tau$ is positive, while $\hat\tau_{12}$ is negative.

\setlength{\tabcolsep}{4pt}
\begin{table}[!h]
  \centering
  \caption{\label{PET-CT-distant}Maximised  ML and CL log-likelihoods, estimates and standard errors (SE)  of the hybrid copula mixed models with different choices of parametric families of copulas and margins for the PET
  modality to diagnose 
  distant 
  lymph node metastasis. }
\vspace{1cm}
    \begin{tabular}{ccccccccccccccc}
    \hline
           & \multicolumn{14}{c}{Normal margins} \\
    \hline
          & \multicolumn{2}{c}{BVN} & \multicolumn{2}{c}{Frank} & \multicolumn{2}{c}{Cln180/270} & \multicolumn{2}{c}{Cln 180/90} & \multicolumn{2}{c}{Cln 0/270} & \multicolumn{2}{c}{Cln 0/90} & \multicolumn{2}{c}{CL} \\
          & Est.  & SE    & Est.  & SE    & Est.  & SE    & Est.  & SE    & Est.  & SE    & Est.  & SE    & Est.  & SE \\\hline
    $\pi_1$ & 0.82  & 0.02  & 0.82  & 0.02  & 0.83  & 0.03  & 0.82  & 0.02  & 0.82  & 0.03  & 0.82  & 0.02  & 0.85  & 0.02 \\
    $\pi_2$ & 0.87  & 0.02  & 0.87  & 0.02  & 0.87  & 0.03  & 0.87  & 0.02  & 0.87  & 0.03  & 0.87  & 0.02  & 0.88  & 0.02 \\
    $\pi_3$ & 0.57  & 0.06  & 0.56  & 0.06  & 0.57  & 0.06  & 0.55  & 0.06  & 0.59  & 0.07  & 0.59  & 0.06  & 0.59  & 0.07 \\
    $\s_1$ & 0.65  & 0.16  & 0.65  & 0.16  & 0.65  & 0.17  & 0.64  & 0.15  & 0.68  & 0.17  & 0.65  & 0.15  & 0.64  & 0.09 \\
    $\s_2$ & 0.94  & 0.21  & 0.91  & 0.20  & 0.94  & 0.23  & 1.01  & 0.22  & 0.93  & 0.23  & 0.99  & 0.21  & 0.89  & 0.10 \\
    $\s_3$ & 0.87  & 0.20  & 0.93  & 0.21  & 0.94  & 0.21  & 0.92  & 0.21  & 0.89  & 0.22  & 0.84  & 0.19  & 0.97  & 0.18 \\
    $\tau_{12}$ & -0.31 & 0.25  & -0.45 & 0.27  & -0.19 & 0.40  & -0.28 & 0.25  & -0.27 & 0.38  & -0.30 & 0.27  & 0.00  & - \\
    $\tau_{13}$ & 0.52  & 0.20  & 0.51  & 0.21  & 0.51  & 0.21  & 0.46  & 0.21  & 0.53  & 0.22  & 0.49  & 0.22  & 0.00  & - \\
    $\tau$ & 0.60  & 0.41  & 0.57  & 0.36  & 0.63  & 0.61  & 0.73  & 0.32  & 0.63  & 0.59  & 0.72  & 0.32  & 0.00  & - \\\hline
    $\log L$ & \multicolumn{2}{c}{-174.50} & \multicolumn{2}{c}{-174.46} & \multicolumn{2}{c}{-175.98} & \multicolumn{2}{c}{-175.03} & \multicolumn{2}{c}{-174.93} & \multicolumn{2}{c}{-173.93} & \multicolumn{2}{c}{-178.80} \\
    \hline
    \end{tabular}
\begin{center}

    \begin{tabular}{ccccccccccccc}
    
          & \multicolumn{12}{c}{Beta margins} \\
    \hline
          & \multicolumn{2}{c}{BVN} & \multicolumn{2}{c}{Frank} & \multicolumn{2}{c}{Cln 180/270} & \multicolumn{2}{c}{Cln 180/90} & \multicolumn{2}{c}{Cln 0/270} & \multicolumn{2}{c}{Cln 0/90} \\
          & Est.  & SE    & Est.  & SE    & Est.  & SE    & Est.  & SE    & Est.  & SE    & Est.  & SE \\\hline
    $\pi_1$ & 0.80  & 0.03  & 0.80  & 0.03  & 0.81  & 0.03  & 0.80  & 0.02  & 0.80  & 0.03  & 0.80  & 0.02 \\
    $\pi_2$ & 0.84  & 0.03  & 0.83  & 0.03  & 0.84  & 0.03  & 0.84  & 0.03  & 0.83  & 0.03  & 0.84  & 0.03 \\
    $\pi_3$ & 0.56  & 0.05  & 0.55  & 0.05  & 0.55  & 0.05  & 0.54  & 0.05  & 0.57  & 0.06  & 0.58  & 0.05 \\
    $\g_1$ & 0.06  & 0.03  & 0.06  & 0.03  & 0.06  & 0.03  & 0.05  & 0.02  & 0.06  & 0.03  & 0.06  & 0.03 \\
    $\g_2$ & 0.10  & 0.04  & 0.09  & 0.04  & 0.10  & 0.04  & 0.11  & 0.04  & 0.10  & 0.04  & 0.10  & 0.04 \\
    $\g_3$ & 0.14  & 0.05  & 0.15  & 0.05  & 0.16  & 0.05  & 0.15  & 0.05  & 0.15  & 0.06  & 0.13  & 0.05 \\
    $\tau_{12}$ & -0.31 & 0.24  & -0.46 & 0.27  & -0.23 & 0.49  & -0.26 & 0.25  & -0.29 & 0.38  & -0.27 & 0.25 \\
    $\tau_{13}$ & 0.53  & 0.20  & 0.53  & 0.21  & 0.51  & 0.21  & 0.47  & 0.21  & 0.55  & 0.22  & 0.50  & 0.22 \\
    $\tau$ & 0.65  & 0.38  & 0.60  & 0.35  & 0.66  & 0.58  & 0.76  & 0.30  & 0.67  & 0.56  & 0.76  & 0.30 \\\hline
    $\log L$ & \multicolumn{2}{c}{-175.76} & \multicolumn{2}{c}{-175.69} & \multicolumn{2}{c}{-177.17} & \multicolumn{2}{c}{-176.37} & \multicolumn{2}{c}{-176.14} & \multicolumn{2}{c}{-175.33} \\
    \hline
    \end{tabular}
    
\end{center}
\begin{footnotesize}
Cln $\alpha/\beta$ denotes a hybrid copula mixed with copula distributions  the Clayton rotated by $\alpha$ and $\beta$ degrees for the $\{C(;\tau),C_{13}(;\tau_{13})\}$ and $C_{12}(;\tau_{12})$ copulas, respectively.
\end{footnotesize}
\end{table}

In all the meta-analyses,   improvement over the 
hybrid copula mixed model composed of  BVN copulas and normal margins, that is the same with the hybrid GLMM  in \cite{chen-etal-2015-jrssc}, has been revealed in terms of the likelihood principle.  
\cite{chen-etal-2015-jrssc}, instead of relying to separate meta-analyses for each type of imaging modality and type of metastasis, analyzed all the data by assuming normal margins for the random effects with equal between-studies variances  
in transformed sensitivity, specificity, and disease prevalence   for different imaging modalities or stages of metastasis.  These assumptions are quite strong and we have shown,  with the subgroup analysis in Tables \ref{US}-\ref{PET-CT-distant}, that are  substantially violated. In fact,   
between study variances  are distinct in  each type  of imaging modality or stage of metastasis and for the US imaging modality even the assumption of normal margins is not valid.

\section{\label{sec-disc}Discussion}
We have proposed a hybrid copula mixed model for  meta-analysis of diagnostic test accuracy studies. It jointly models the disease prevalence along with diagnostic test sensitivity and specificity in cohort studies, and sensitivity and specificity in case-control studies. 
Our general model includes the hybrid GLMM \citep{chen-etal-2015-jrssc} as a special case  and  can provide an improvement over the latter based on log-likelihood.  Hence, a better statistical inference for the meta-analytic parameters and their between-study variances is achieved. Nevertheless the meta-analytic parameters are a univariate inference, and hence it is the univariate marginal distribution that  matters the most and not the type of the copula. The proposed hybrid copula mixed model  calls on both normal and beta univariate margins and thus can  operate on the transformed and original scale of sensitivity, specificity and disease prevalence, respectively.

Though typically the focus of meta-analysis has been to derive the  summary-effect estimates, there is increasing interest in drawing predictive inference. 
In fact, if the interest is only to overall sensitivity, specificity and prevalence  then the overall test accuracy across studies will not be clearly defined.  Different studies use different thresholds for a positive test result, thus  the overall summary-effect estimates do not make sense. 
Instead, some form of SROC curve makes much more sense and will help decision makers to assess the actual diagnostic accuracy of a diagnostic test \citep{Nikoloulopoulos2016-letter-smmr}. 
SROC curves are deduced for our model  through the quantile regression techniques developed by \cite{Nikoloulopoulos2015b}.
For the hybrid copula mixed model, the model parameters (including dependence parameters), the choice of the copula, and the choice of the margin affect the shape of the SROC curve.   
Among the parametric families of copulas in Table \ref{2fam} the tail dependence varies, and is a property to consider when choosing amongst different families of copulas, and, hence affects the shape of SROC curves, i.e., prediction.  SROC  will essentially show the
effect of different  model (random effect distribution) assumptions, since it is an inference that depends on the joint distribution \citep{Nikoloulopoulos2015b}. 
Given that the  CL estimation assumes independence among the random effects, it provides  identical fit for any copula mixed model, since all the parametric families of copulas in Table \ref{2fam} contain the independence copula as a special case.
Hence,  the big limitation of the CL method is that it cannot be used to produce the SROC curves, since the  dependence parameters  affect the shape of the SROC curve and these are set to independence by definition.

It has been reported in the literature that in the trivariate  GLMM \citep{chu-etal-2009} and hybrid GLMM \citep{chen-etal-2015-jrssc} estimation problems     relating to the correlation parameters exist, such as non-convergence. Here instead of a trivariate normal distribution we use a vine copula distribution,  and  in particular  a truncated at level-1 vine copula (conditional independence), which allows both parsimony and flexible (tail) dependence. In fact, we propose a numerically stable ML estimation technique based on Gauss-Legendre quadrature; the crucial step is to convert from independent  to dependent quadrature points.   
However, the additional feature of having to estimate the associations among the random effects has been found to require larger sample sizes than in CL estimation where these parameters are set to independence. The application example includes cases with an adequate number of  individual studies per study design. For meta-analyses with fewer (especially cohort) studies  the bivariate copula mixed model to obtain estimates of diagnostic sensitivity and specificity but not prevalence should be fitted instead.  Future research will deal with the development of  penalized likelihood methods for optimising inference about the association parameters of the hybrid copula mixed model when the number of available study summaries is small.

We also plan to provide extensions of the model to account for partial verification bias. This is a feature that has been already developed for the hybrid GLMM \citep{ma-etal-2014-smmr}.

\section*{\label{software}Software}
{\tt R} functions to  implement the hybrid vine copula mixed model for meta-analysis of diagnostic test accuracy case-contol and cohort studies are part of the {\tt  R} package {\tt  CopulaREMADA} \citep{Nikoloulopoulos-2015}.  

\section*{Acknowledgement}
We would like to thank Dr  Yong Chen (University of Texas) and Professor Haitao Chu  (University of Minnesota) for providing the melanoma data.


\begin{thebibliography}{}
\itemsep=0pt
\bibitem[Aas et~al., 2009]{aasetal09}
Aas, K., Czado, C., Frigessi, A., and Bakken, H. (2009).
\newblock Pair-copula constructions of multiple dependence.
\newblock {\em Insurance: {M}athematics \& {E}conomics}, 44:182--198.

\bibitem[Brenner and Gefeller, 1997]{brenner-gefeller-1997}
Brenner, H. and Gefeller, O. (1997).
\newblock Variation of sensitivity, specificity, likelihood ratios and
  predictive values with disease prevalence.
\newblock {\em Statistics in Medicine}, 16(9):981--991.

\bibitem[Chen et~al., 2015]{chen-etal-2015-jrssc}
Chen, Y., Liu, Y., Ning, J., Cormier, J., and Chu, H. (2015).
\newblock A hybrid model for combining case–control and cohort studies in
  systematic reviews of diagnostic tests.
\newblock {\em Journal of the Royal Statistical Society: Series C (Applied
  Statistics)}, 64(3):469--489.

\bibitem[Chen et~al., 2014]{Chen-etal-smmr-2014}
Chen, Y., Liu, Y., Ning, J., Nie, L., Zhu, H., and Chu, H. (2014).
\newblock A composite likelihood method for bivariate meta-analysis in
  diagnostic systematic reviews.
\newblock {\em Statistical Methods in Medical Research}.
\newblock
  \href{http://dx.doi.org/10.1177/0962280214562146}{DOI:10.1177/0962280214562146}.

\bibitem[Chu and Cole, 2006]{Chu&Cole2006}
Chu, H. and Cole, S.~R. (2006).
\newblock Bivariate meta-analysis of sensitivity and specificity with sparse
  data: a generalized linear mixed model approach.
\newblock {\em Journal of Clinical Epidemiology}, 59(12):1331--1332.

\bibitem[Chu et~al., 2009]{chu-etal-2009}
Chu, H., Nie, L., Cole, S.~R., and Poole, C. (2009).
\newblock Meta-analysis of diagnostic accuracy studies accounting for disease
  prevalence: Alternative parameterizations and model selection.
\newblock {\em Statistics in Medicine}, 28(18):2384--2399.

\bibitem[Demidenko, 2004]{Demidenko04}
Demidenko, E. (2004).
\newblock {\em Mixed {M}odels: {T}heory and {A}pplications}.
\newblock John Wiley \& Sons, Hoboken, New Jersey.

\bibitem[Jerant et~al., 2000]{Jerant-etal-2000}
Jerant, A., Johnson, J., Sheridan, C., and Caffrey, T. (2000).
\newblock Early detection and treatment of skin cancer.
\newblock {\em American Family Physician}, 62(2):357--368+375--376+381--382.

\bibitem[Joe, 1997]{joe97}
Joe, H. (1997).
\newblock {\em Multivariate {M}odels and {D}ependence {C}oncepts}.
\newblock Chapman \& Hall, London.

\bibitem[Joe, 2011]{joe2010a}
Joe, H. (2011).
\newblock Dependence comparisons of vine copulae with four or more variables.
\newblock In Kurowicka, D. and Joe, H., editors, {\em Dependence Modeling:
  Handbook on Vine Copulae}, chapter~7, pages 139--164. World Scientific,
  Singapore.

\bibitem[Joe, 2014]{joe2014}
Joe, H. (2014).
\newblock {\em Dependence Modeling with Copulas}.
\newblock Chapman \& Hall, London.

\bibitem[Joe et~al., 2010]{joeetal10}
Joe, H., Li, H., and Nikoloulopoulos, A.~K. (2010).
\newblock Tail dependence functions and vine copulas.
\newblock {\em Journal of Multivariate Analysis}, 101:252--270.

\bibitem[Kurowicka and Joe, 2011]{Kurowicka-Joe-2011}
Kurowicka, D. and Joe, H. (2011).
\newblock {\em Dependence {M}odeling -- {H}andbook on {V}ine {C}opulae}.
\newblock World Scientific Publishing Co, Singapore.

\bibitem[Leeflang et~al., 2013]{Leeflang-etal-2013}
Leeflang, M.~M., Rutjes, A.~W., Reitsma, J.~B., Hooft, L., and Bossuyt, P.~M.
  (2013).
\newblock Variation of a test's sensitivity and specificity with disease
  prevalence.
\newblock {\em Canadian Medical Association Journal}, 185(11):E537--E544.

\bibitem[Leeflang et~al., 2009]{leeflang-etal-2009}
Leeflang, M. M.~G., Bossuyt, P. M.~M., and Irwig, L. (2009).
\newblock Diagnostic test accuracy may vary with prevalence: implications for
  evidence-based diagnosis.
\newblock {\em Journal of Clinical Epidemiology}, 62(1):5--12.

\bibitem[Ma et~al., 2014]{ma-etal-2014-smmr}
Ma, X., Chen, Y., Cole, S.~R., and Chu, H. (2014).
\newblock A hybrid bayesian hierarchical model combining cohort and
  case--control studies for meta-analysis of diagnostic tests: Accounting for
  partial verification bias.
\newblock {\em Statistical methods in medical research}.
\newblock
  \href{http://dx.doi.org/10.1177/0962280214536703}{DOI:10.1177/0962280214536703}.

\bibitem[Nash, 1990]{nash90}
Nash, J. (1990).
\newblock {\em Compact Numerical Methods for Computers: Linear Algebra and
  Function Minimisation}.
\newblock Hilger, New York.
\newblock 2nd edition.

\bibitem[Nelsen, 2006]{nelsen06}
Nelsen, R.~B. (2006).
\newblock {\em An {I}ntroduction to {C}opulas}.
\newblock Springer-Verlag, New York.

\bibitem[Nikoloulopoulos, 2015a]{Nikoloulopoulos2015b}
Nikoloulopoulos, A.~K. (2015a).
\newblock A mixed effect model for bivariate meta-analysis of diagnostic test
  accuracy studies using a copula representation of the random effects
  distribution.
\newblock {\em Statistics in Medicine}, 34:3842--3865.

\bibitem[Nikoloulopoulos, 2015b]{Nikoloulopoulos2015c}
Nikoloulopoulos, A.~K. (2015b).
\newblock A vine copula mixed effect model for trivariate meta-analysis of
  diagnostic test accuracy studies accounting for disease prevalence.
\newblock {\em Statistical Methods in Medical Research}.
\newblock
  \href{http://dx.doi.org/10.1177/0962280215596769}{DOI:10.1177/0962280215596769}.

\bibitem[Nikoloulopoulos, 2016a]{Nikoloulopoulos2016-letter-smmr}
Nikoloulopoulos, A.~K. (2016a).
\newblock Comment on `{A} vine copula mixed effect model for trivariate
  meta-analysis of diagnostic test accuracy studies accounting for disease
  prevalence'.
\newblock {\em Statistical Methods in Medical Research}, 25(2):988--991.

\bibitem[Nikoloulopoulos, 2016b]{Nikoloulopoulos-2015}
Nikoloulopoulos, A.~K. (2016b).
\newblock {\em {CopulaREMADA}: {C}opula mixed effect models for bivariate and
  trivariate meta-analysis of diagnostic test accuracy studies}.
\newblock {R} package version 1.0.

\bibitem[Nikoloulopoulos and Joe, 2015]{nikoloulopoulos&joe12}
Nikoloulopoulos, A.~K. and Joe, H. (2015).
\newblock Factor copula models for item response data.
\newblock {\em Psychometrika}, 80:126--150.

\bibitem[Nikoloulopoulos and Karlis, 2008]{Nikoloulopoulos&karlis08CSDA}
Nikoloulopoulos, A.~K. and Karlis, D. (2008).
\newblock Copula model evaluation based on parametric bootstrap.
\newblock {\em Computational Statistics \& Data Analysis}, 52:3342--3353.

\bibitem[Sklar, 1959]{sklar1959}
Sklar, M. (1959).
\newblock Fonctions de r\'epartition \`a {$n$} dimensions et leurs marges.
\newblock {\em Publications de l'Institut de Statistique de l'Universit\'e de
  Paris}, 8:229--231.

\bibitem[Stober et~al., 2013]{Stober-joe-czado2013}
Stober, J., Joe, H., and Czado, C. (2013).
\newblock Simplified pair copula constructions-limitations and extensions.
\newblock {\em Journal of Multivariate Analysis}, 119:101--118.

\bibitem[Stroud and Secrest, 1966]{Stroud&Secrest1966}
Stroud, A.~H. and Secrest, D. (1966).
\newblock {\em Gaussian Quadrature Formulas}.
\newblock Prentice-Hall, Englewood Cliffs, NJ.

\bibitem[Varin, 2008]{varin08}
Varin, C. ({2008}).
\newblock {On composite marginal likelihoods}.
\newblock {\em {Advances in Statistical Analysis}}, {92}:{1--28}.

\bibitem[Varin et~al., 2011]{Varin-etal2011}
Varin, C., Reid, N., and Firth, D. (2011).
\newblock An overview of composite likelihood methods.
\newblock {\em Statistica Sinica}, 21:5--42.

\bibitem[Xing et~al., 2011]{Xing-etal-2011}
Xing, Y., Bronstein, Y., Ross, M.~I., Askew, R.~L., Lee, J.~E., Gershenwald,
  J.~E., Royal, R., and Cormier, J.~N. (2011).
\newblock Contemporary diagnostic imaging modalities for the staging and
  surveillance of melanoma patients: a meta-analysis.
\newblock {\em Journal of the National Cancer Institute}, 103(2):129--142.

\end{thebibliography}
\end{document}